\documentclass[aps, pre, twocolumn, 10pt, superscriptaddress, showpacs, nofootinbib, floatfix, groupedaddress]{revtex4-2}
\usepackage{graphicx}
\usepackage{xcolor}
\usepackage{dcolumn}
\usepackage{amsmath}
\usepackage{bm}
\usepackage{latexsym}
\usepackage{mathrsfs}
\usepackage{hyperref}
\hypersetup{
	colorlinks = true,
	linkcolor  = red,
	filecolor  = blue,
	urlcolor   = magenta,
}
\usepackage[normalem]{ulem}

\usepackage{pdfpages} 
\usepackage{pgffor} 

\makeatletter
\AtBeginDocument{\let\LS@rot\@undefined}
\makeatother

\def\supplementfilename{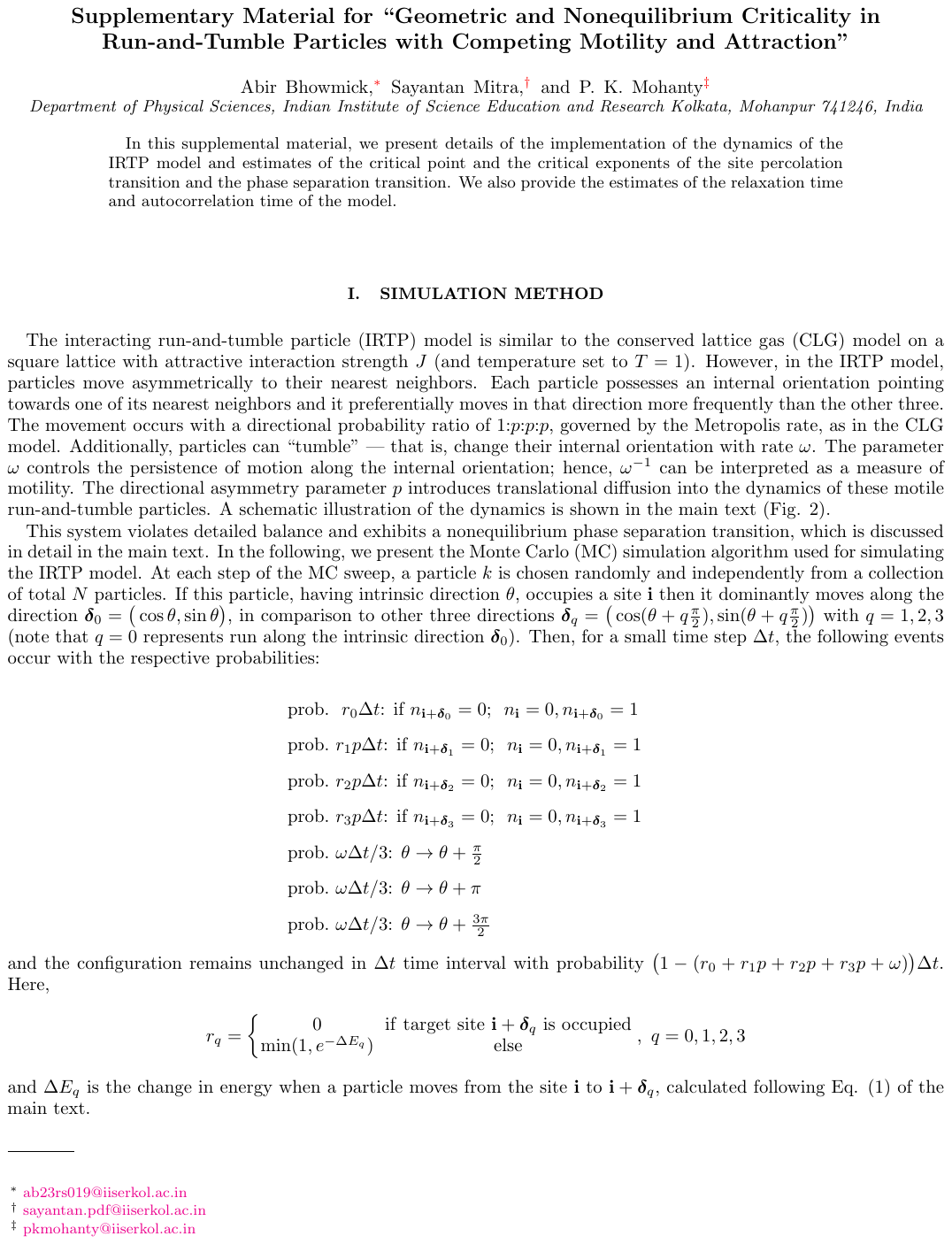}

\pdfximage{\supplementfilename}
\def\numbersupplementpages{\the\pdflastximagepages}

\newif\ifarXiv
\arXivtrue 

\newcommand{\pst}{phase separation transition}
\newcommand{\pss}{phase-separated state}
\newcommand{\iuc}{IUC}
\newcommand{\m}{MIPS}
\newcommand{\cg}{CLG}
\newcommand{\rtp}{RTPs}
\newcommand{\smx}{s_{\text{max}}}
\newcommand{\bn}{n_{\bf i}}
\newcommand{\be}{\begin{equation}} 
	\newcommand{\ee}{\end{equation}} 
\newcommand{\bea}{\begin{eqnarray}} 
	\newcommand{\eea}{\end{eqnarray}}

\newcommand{\wc}{\omega_{c}}
\newcommand{\Jc}{J_{c}}
\newcommand{\Jcs}{J_{c}^*}
\newcommand{\rc}{\rho_{c}}
\newcommand{\nuI}{\nu^{\text{I}}}
\newcommand{\btI}{\beta^{\text{I}}}
\newcommand{\gmI}{\gamma^{\text{I}}}
\newcommand{\nuM}{\nu^{M}}
\newcommand{\btM}{\beta^{M}}
\newcommand{\gmM}{\gamma^{M}}
\newcommand{\el}{\ell}
\newcommand{\ro}{\rho}
\newcommand{\w}{\omega}
\newcommand{\rp}{\rho_+}
\newcommand{\rn}{\rho_-}
\newcommand{\dn}{\Delta N}

\newcommand{\tp}{\tilde{\phi}}
\newcommand{\tc}{\tilde{\chi}}
\newcommand{\avg}[1]{\langle #1 \rangle}

\begin{document}
	
	\title{Geometric and Nonequilibrium Criticality in Run-and-Tumble Particles with Competing Motility and Attraction}
	
	\author{Abir Bhowmick}
	\email{ab23rs019@iiserkol.ac.in}
	\author{Sayantan Mitra}
	\email{sayantan.pdf@iiserkol.ac.in }
	\author{P. K. Mohanty}
	\email{pkmohanty@iiserkol.ac.in}
	\affiliation {Department of Physical Sciences, Indian Institute of Science Education and Research Kolkata, Mohanpur 741246, India}
	
	\begin{abstract}
		Self-propulsion in run-and-tumble particles (RTPs) generates effective attractive interactions that can drive motility-induced phase separation (MIPS), a phenomenon absent in passive systems. Here, we investigate RTPs in the presence of explicit attractive interactions and show that, at high motility, such interactions can suppress MIPS, yielding a homogeneous phase. Upon further increasing the attraction strength, phase separation reappears, giving rise to a re-entrant transition. We characterize this transition by analyzing the percolation properties of dense clusters, which provide geometric signatures of phase separation. Along the resulting critical line, we find continuously varying critical exponents, while certain scaling functions remain unchanged and coincide with those of equilibrium lattice gas models undergoing interacting percolation, which is in the  Ising-percolation universality class. These results reveal that the MIPS transition in interacting RTP systems  exhibit Ising superuniversality, thereby establishing a connection between nonequilibrium active matter and classical critical behavior.
	\end{abstract}
	
	\maketitle
	
	\section{Introduction\label{introduxtion}}
	
	Active matter is a special class of nonequilibrium systems where the constituents self-propel by consuming energy from the environment \cite{Ramaswamy2010, Cates2012, Marchetti2013}. A vast range of systems composed of motile or self-propelled particles falls within this class. Some examples include flocks of birds \cite{Ballerini2008}, fish schools \cite{Katz2011}, actin filaments \cite{Schaller2010}, and microtubules \cite{Sumino2012} in the context of the study of locomotion and colloidal systems \cite{Theurkauff2012, Buttinoni2013, Palacci2013, Bricard2013}. Two of the simplest and most widely studied models of active particle dynamics are the \textit{run-and-tumble particles} (\rtp)~\cite{Schnitzer1993, H.C.Berg2004} and the \textit{active Brownian particles} (ABPs) \cite{Romanczuk2012}. In the case of ABPs, the direction of motion changes smoothly via rotational diffusion at each time step, whereas \rtp~execute a period of persistent motion along their internal orientation called a ``run,'' followed by a ``tumble'' where they change their internal orientation with some rate. Even this minimal model is capable of explaining many complex systems, such as the motion of \textit{Escherichia coli}; studied extensively in several works \cite{H.C.Berg2004, Polin2009}.
	
	Among the many fascinating characteristics shown by active particle systems, such as nonequilibrium steady-states \cite{Cates2013, Tailleur2009, Enculescu2011, Lee2013, Romanczuk2012}, clustering \cite{Slowman2016, Slowman2017}, ratchet effects \cite{Reichhardt2017}; the most intriguing one that draws the attention of many researchers is the motility-induced phase separation (\m)~\cite{Tailleur2008,  Redner2013, Redner2013a, Stenhammar2013, Cates2015, Patch2017} where the particles, even in the absence of any attractive interaction, form high-density clusters well separated from a low-density region. In fact, the speed of active particles is comparatively lower in the regions where the local density is higher; this, in effect, increases the local density further, making a uniformly mixed suspension unstable. This instability gives rise to motility-induced phase separation, where a dilute, motile gas phase coexists with a less motile dense liquid phase. A MIPS transition, arising in the absence of any attractive interactions, is one of the most fascinating phenomena observed in active-matter systems. While most of the studies are based on numerical simulations, there are a few theoretical studies that employ hydrodynamic analysis \cite{Fily2012, Bialk2013, Marchetti2013}, some use agent-based modeling \cite{Schweitzer2018, Ziepke2022}, and also there are lattice-based approaches \cite{Thompson2011, Slowman2016, Slowman2017, Mallmin2019, Dandekar2020, Ray2024}. In one dimension (1D), even though hydrodynamic analysis supports the existence of \m \cite{KourbaneHoussene2018, Seplveda2016}, a recent lattice-based study cast doubt on it \cite{Mukherjee2023a}. However, in two dimensions (2D), numerous on-lattice \cite{Dittrich2021, Soto2014, Whitelam2018, Yao2025} and off-lattice \cite{Solon2015, Siebert2018} models exhibit \m~transition. 
	
	Persistent motion of active particles is essential for phase separation, but it alone does not guarantee it. A recent study \cite{Ray2024} demonstrates that passive particles undergoing phase separation due to interparticle attraction $J$, as in a lattice gas model, require significantly stronger attraction to phase separate once motility is introduced. In fact, no \m~phase was observed when the attraction was absent ($J=0$). A later on-lattice study of RTPs with zero attraction $(J=0)$ \cite{Soumya2024}, showed that translational diffusion, which allows particles to move beyond their primary propulsion direction, plays a crucial role in enabling the \m~transition. By examining the relation between percolation phenomena and phase separation, the authors suggested that the \m~transition may belong to the Ising superuniversality class, characterized by continuously varying critical exponents alongside some scaling functions similar to those of the equilibrium lattice gas phase separation. This interpretation aligns with several other studies \cite{Ray2024, Partridge2019, Maggi2021, Dittrich2021} proposing that \m~transitions fall within the $2$D Ising universality class (\iuc).
	
	It is well established that active particles can generate effective pairwise attraction due to their persistent self-propelled motion \cite{Cates2013, Redner2013, Redner2013a}, which drives their clustering behavior. One might intuitively expect that adding explicit attractive interactions would enhance phase separation. However, numerical studies suggest that this is not always the case \cite{Redner2013a, Ray2024}. A recent analytical work \cite{Urna2025} further demonstrates that, in systems of ABPs with speed inhomogeneity, explicit attraction can actually induce an effective short-range repulsion. While several studies have explored the interplay between activity and attraction \cite{Redner2013a, Caprini2023, Chakraborti2024, Ray2024, Caprini2023, Sota2025}, the precise role of attractive interactions in modulating \m~remains not fully understood.
	
	In this article, we investigate the effect of attractive interaction strength $J$, on a system of hard-core \rtp~on a square lattice, focusing on its impact on \m~transition. The model we study, referred to as the interacting run-and-tumble particle (IRTP) model, consists of \rtp~that possess an internal orientation pointing toward one of the four nearest-neighbor directions. During a ``run'' event, a particle attempts to move to a neighboring site, with a higher rate in the direction of its internal orientation. In a ``tumble'' event, the internal orientation is reassigned randomly to one of the other three directions. The event of tumbling occurs at a rate $\w$, and its inverse $\w^{-1}$, which defines the persistence time, serves as a quantitative measure of particle motility.
	
	We find that in the high motility regime, the phase-separated state observed at $J=0$ is disrupted as attractive interactions are introduced. However, upon increasing $J$ further, the system re-enters a phase-separated state, giving rise to a re-entrant \m~transition. Along the critical line in the $\w$--$J$ plane, the critical exponents are observed to vary continuously, signaling nonuniversal behavior. Nevertheless, the transition belongs to the Ising superuniversality class in the sense that, although the exponents change continuously, some of the scaling functions coincide with those of the standard equilibrium phase transition in the two-dimensional lattice gas model.
	
	The article is organized as follows: In Sec.~\ref{sec: II}, we briefly describe the model along with the limiting cases of its parameters. In Sec.~\ref{sec: III}, we discuss the choice of order parameters and the associated challenges in systematically studying both the geometric transition (percolation) and the underlying phase separation transition. Section~\ref{sec: IV} presents the numerical methods and results related to the percolation and MIPS transitions. In Sec.~\ref{sec: V}, we analyze the role of density—especially in systems lacking particle-hole symmetry (such as 2D lattice gases or Ising models)—in determining the critical parameters and the corresponding static exponents. In Sec.~\ref{sec: VI}, we argue that the phase separation transition of interacting \rtp~belongs to the Ising or $Z_2$ superuniversality class. Finally, in Sec.~\ref{sec: VII}, we discuss key implications of our findings, and conclude in Sec.~\ref{sec: VIII}.

	\section{The Model\label{sec: II}}
	
	\begin{figure}[t]
		\centering
		\includegraphics[width = \linewidth]{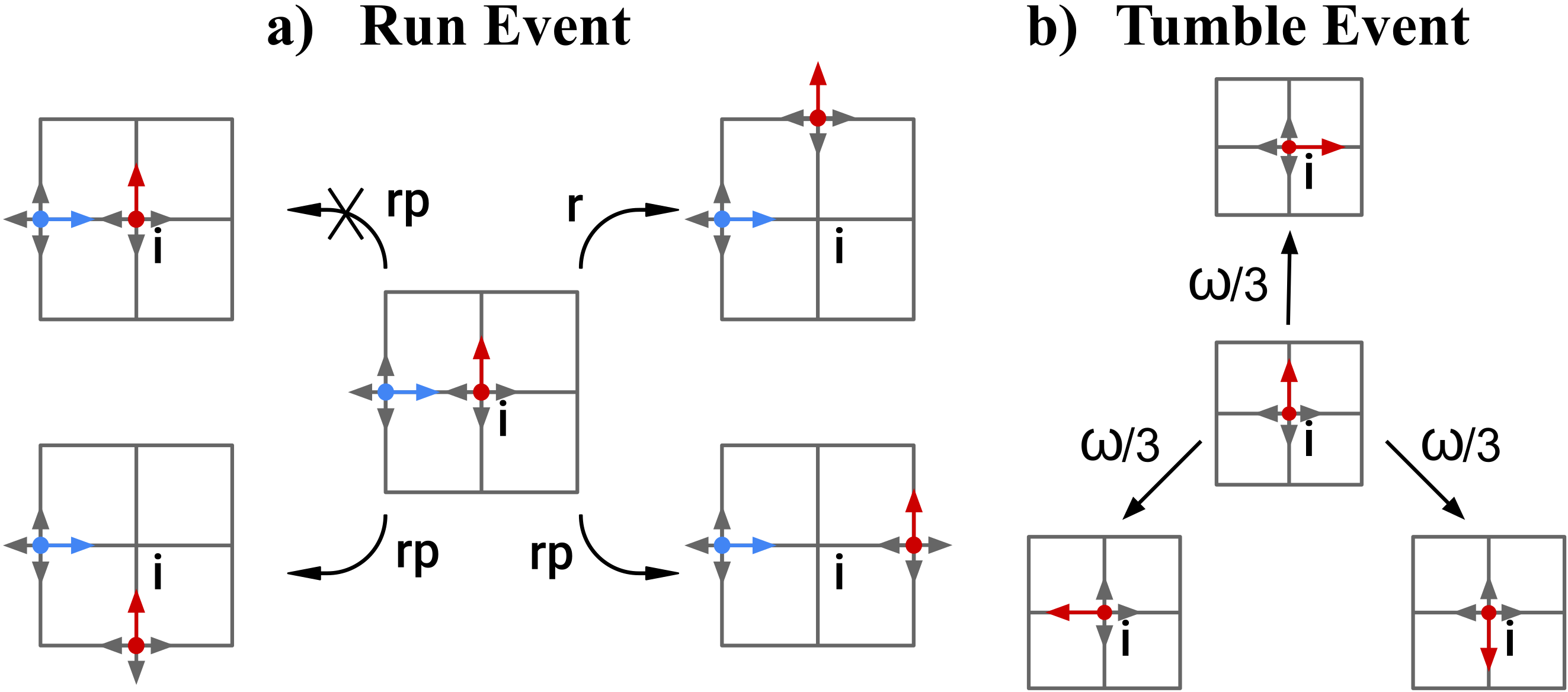}
		\caption{(Color online) Dynamics of IRTP model: An RTP (the central particle, marked red), chosen out of $N$, with site index $\mathbf{i}$ and internal orientation  $\theta_k= \frac\pi2$ (pointing up) can (a) run or (b) tumble with indicated rates. Here $r={\rm min} (1, e^{-\Delta E})$ is the Metropolis rate with respect to Eq.~\eqref{eq: E}, the parameter $0\le p<1$ generates translational diffusion, and $\w>0$ is a constant tumble rate. Note that the hardcore nature of \rtp~does not allow the particle at ${\bf i}$ to move to its left, as this site is already occupied by another RTP.}
		\label{fig:lattice_dynamics_pic}
	\end{figure}
	We consider $N$ \rtp~on a square lattice $\cal L$ of $L^2$ sites labeled by $\mathbf{i} \equiv (x,y)$, with $x, y = 1, 2, \dots, L$ and use periodic boundary conditions in both $x$ and $y$ directions. Each site $\mathbf{i}$ has an occupation number $\bn=\{0,1\}$ that represents whether the site is vacant or filled up. Due to the hardcore (excluded volume) interaction, each site can be occupied by at most one RTP, imposing a constraint on the total number of particles: $\sum_{\mathbf{i}}\bn=N$. Each RTP, labeled as $k=1,2,\dots, N$ has an intrinsic orientation $\theta_k \in  \{0, \frac\pi2, \pi, \frac{3\pi}2\}$ that determines the direction of dominant movement to the nearest neighbor in $\theta_k$ direction, i.e., from the site $\mathbf{i}$ to the direction ${\bf i} + \bm{\delta}_k$, where $\bm{\delta}_k= \left( \cos(\theta_k), \sin(\theta_k) \right)$. Note, that $\mathbf{i} + \bm{\delta}_q$ with $\bm{\delta}_q=\left( \cos(q\frac\pi2), \sin(q\frac\pi2) \right);~~q=0,1,2,3$, are the four nearest neighbors of a site ${\bf i} \in {\cal L}.$ 
	
	Besides the hardcore interaction, the \rtp~experience a nearest-neighbor attraction, described by an energy function
	\begin{equation}
		\label{eq: E}
		E = -J \sum_{{\bf i} \in {\cal L} } {\sum_{q=0}^3} \bn n_{\mathbf{i} + \bm{\delta}_q},
	\end{equation}
	where $J>0$. Due to the presence of attractive interaction among the \rtp, we refer to this model as the IRTP model. Note that the internal orientation of \rtp~do not contribute to energy; they only dictate their direction of motion. During a ``run'' event, an RTP at the site ${\bf i}$ moves to one of its neighbors with a Metropolis rate $r= {\rm min}\{1, e^{-\Delta E}\}$, where $\Delta E$ is the difference in energy between final and initial configurations. However, the selection of neighbors is asymmetric, with a ratio of 1:3$p$ between the internal orientation and the remaining three spatial directions. This parameter $0\le p\le1$ accommodates the translational diffusion of \rtp. In addition to ``run,'' an RTP can ``tumble'' with rate $\w$ and change its internal orientation to any of the other three lattice directions, chosen randomly. A schematic picture of the run and tumble events is described in Fig.~\ref{fig:lattice_dynamics_pic}.
	
	We intend to study the phase separation transition of this IRTP model using Monte Carlo simulations, which will be discussed in the subsequent sections. But before that, let us look at some interesting limits.
	
	\textit{$p=1$ case}. For $p=1$, \rtp~choose any of the four directions with equal probability. Then, the internal orientation is no more special, and tumbling loses its meaning. The dynamics then become identical to that of the conserved lattice gas (\cg) model in 2D with the temperature set at $T=1$. Thus, like 2D \cg~model, one would expect the IRTP system to exhibit an equilibrium phase separation transition at the Onsager value (in units of $k_B=1$ and $T=1$)\cite{Onsager1944} 
	\begin{equation}
		\label{eq: Jcs}
		\Jcs = 2 \ln(1 + \sqrt 2),~\rc = 1/2,
	\end{equation}
	where the coexistence disappears (i.e., where the gas density is so high that it cannot be distinguished from the liquid density). The critical nature of this equilibrium transition is characterized by the Ising critical exponents associated with the order parameter $\phi,$ the susceptibility $\chi$, and the correlation length $\xi$ as 
	\begin{equation}
		\label{eq: CritExp_Ising}
		\phi \sim|J-\Jc|^{\btI};~\chi \sim|J-\Jc|^{-\gmI};~\xi \sim|J-\Jc|^{-\nuI}
	\end{equation}
	with $\btI=1/8,~\gmI= 7/4$, and $\nuI=1$. The superscript ``I'' in exponents stands as a reminder of the fact that the transition is in the \iuc.
	
	\textit{$p=0$ case}. For $p=0$ the IRTP model reduces to what has been studied earlier in Ref.~\cite{Ray2024}. In this case, particles can run only along their internal orientation with the Metropolis rate. In the $\w \to \infty$ limit, the internal orientation of particles changes too often between two consecutive run events; in effect, the dynamics of the model reduce to that of the \cg~model, leading to a \pst~at $J=\Jcs$ [Eq.~\eqref{eq: Jcs}]. For any finite $\w$, the \pst~occurs at a critical interaction $\Jc(\w)$, which is larger than $\Jcs$. This indicates that in the absence of any translational diffusion ($p = 0$), increased motility ($\w^{-1}$) hinders cluster formation and the system requires stronger attraction among particles to phase separate.
	
	\textit{$J=0$ case}. In the absence of any interaction, our model closely resembles the model studied recently in Ref.~\cite{Soumya2024}, with the only difference that tumbling of internal orientation $\theta \to \theta +\pi$, was absent there. Using a mapping between phase separation and percolation, the authors found a re-entrant \m~transition in the $\w$-$p$ plane: The system phase separates at intermediate $p$ but remains mixed at both low and high $p$. These findings align well with the absence of \m~at $p=0, J=0$ studied in Ref. \cite{Ray2024} and the lack of transition at $J=0, p=1$, corresponding to the 2D symmetric exclusion process.
	
	The IRTP system we study has both translational diffusion and attractive interaction which serves as an essential foundation for capturing the interplay between \m~transition and conventional phase separation. It has been suggested that the \m~transition in \rtp~belongs to the \iuc~\cite{Partridge2019}. Incorporating attraction among particles brings this model closer in spirit to many off-lattice active-matter systems, where interparticle interactions play a key role in phase separation \cite{Siebert2018, Gonnella2015, MartinRoca2021}. This framework also provides a natural bridge between \m~and the phase separation behavior observed in 2D \cg~models. Through our work, we focus on the geometric aspects of the IRTP system by studying the percolation transition of \rtp~and utilize the known mapping between percolation and phase separation to extract the critical exponents associated with the underlying \m~transition in the presence of attractive interactions.
	
	\section{Relation between  phase separation and  percolation transitions\label{sec: III}}
	
	In the \pss, the density of the system is not homogeneous and the coarse-grained density profile breaks translational symmetry, which is observed as an emergent instability in hydrodynamic theory \cite{ Tailleur2008, Cates2015}. In this phase, the system separates into coexisting high- and low-density regions with densities $\rp$ (liquid) and $\rn$ (gas) respectively, while the conserved particle density $\ro = N/L^2$ is absent locally throughout the system. In the mixed phase, the mean local density of the system is expected to be equal to this density $\ro$ with the system having a translationally invariant coarse-grained density profile.
	
	In the case of 2D \cg~model, where particles interact via the nearest neighbor attraction similar to that of Eq.~\eqref{eq: E}, it is well known that the system exhibits a \pss~when the interaction strength $J$ is increased above   the threshold value $\Jcs$ [Eq.~\eqref{eq: Jcs}]. For $J>\Jcs$ the system co-exists in two different phases having densities $\rp$ and $\rn$, forms a coexistence line in the $\ro$-$J$ plane with its  minima  at $(\rc,~\Jcs)$.  The critical density  in CLG  is  known to be   $\rc=1/2,$ from  the particle-hole symmetry. Below this critical point, the coexistence ceases to exist, and the system becomes homogeneous. Usually   one   considers $\avg{\rp} -\avg{\rn}$ as the order parameter of the \pst. But  a numerical determination of  the exact value of the critical density where  the coexistence lines meet is  usually  difficult  because  the curvature of the coexistence line at $\ro \simeq \rc$ is  very low and the change in the properties of the system is not significant unless one changes the density appreciably. The absence particle-hole symmetry  in active particle systems makes the determination of $\rc$ and hence the associated  $\Jc$  harder. To overcome this challenge, we follow the method proposed in Refs. \cite{Rovere1993, Siebert2018, Partridge2019}. Later, we demonstrate that $\rc$ is very close to $1/2$ and show that the critical points and the static exponents calculated at that density are the same within error limits to those obtained by considering $\rc=1/2$. This insensitive nature of $\rc$ has been indicated earlier in several studies \cite{Rovere1993, Siebert2018, Partridge2019, Soumya2024}.
	
\begin{figure}[t]
		\centering
		\includegraphics[width=0.95\linewidth]{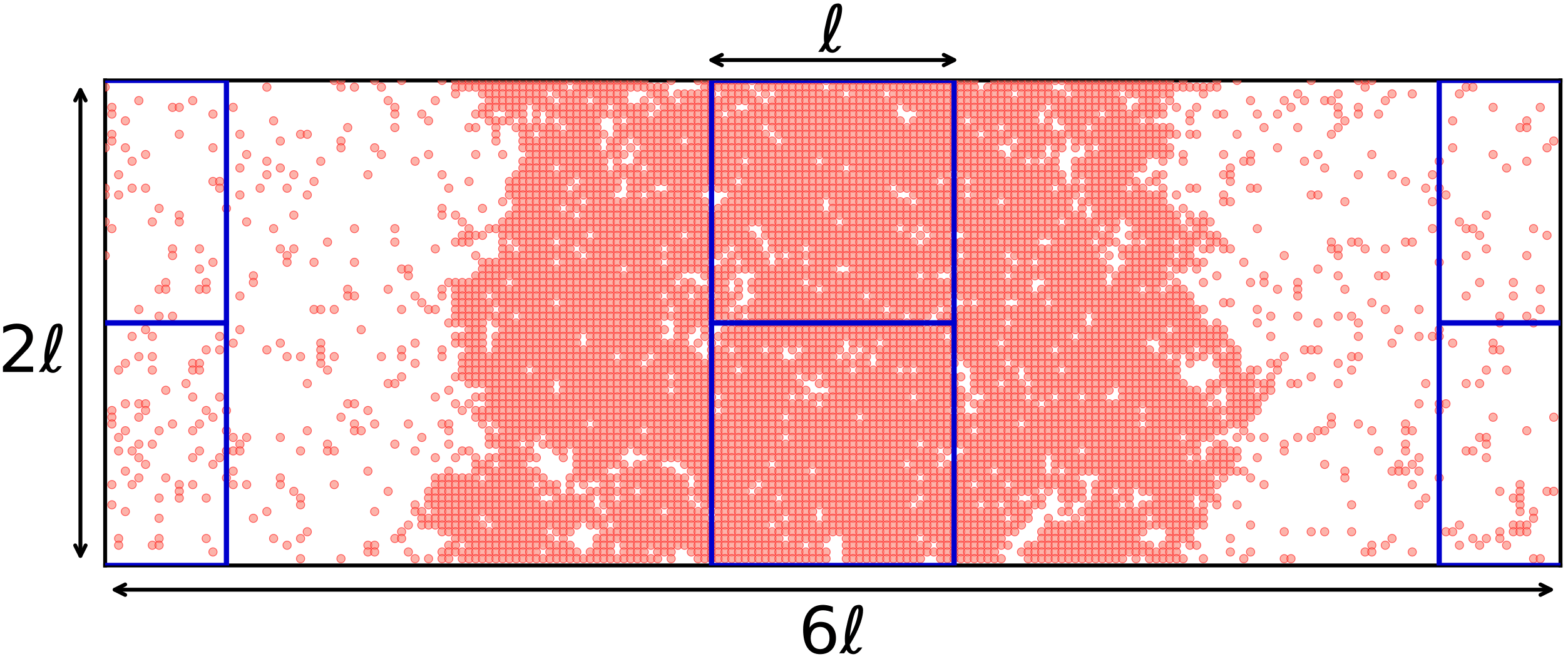}
		\caption{A snapshot of the steady-state of the IRTP system for $\w=0.015$ and $J=0.0$, on a $6\ell\times2\ell$ simulation box (with periodic boundary conditions on both directions) with $\ell=36$. The four square sub-boxes of size  $\ell\times \ell$ are placed in dense and dilute regions to determine coexisting densities $\rp$ (liquid) and  $\rn$ (gas), respectively.}
		\label{fig:ss_MIPS}
	\end{figure}

	Another challenge in studying the phase separation transition of active particles is identifying a suitable order parameter. A common approach involves analyzing the density histogram, which evolves from a single-peaked to a double-peaked profile as the system undergoes phase separation. Another widely used method is to compute the power spectrum of density fluctuations, which exhibits a power-law divergence in the zero-momentum limit at the critical point. Although both these approaches effectively distinguish a phase-separated state from a homogeneous (mixed) one, they suffer from strong finite-size effects, which limit their accuracy in determining the critical point. To address this and determine the critical point more accurately in active-matter systems, Binder and coworkers proposed the \textit{sub-box method} in a series of studies \cite{Binder1981, Rovere1993, Siebert2018}, which is well suited for implementation on a rectangular lattice. The idea is to simulate the system in a $6\el \times 2\el$ rectangular box containing $N$ particles and analyze four $\el \times \el$ sub-boxes in each steady-state configuration:  the first two, referred to as \textit{dense boxes} are centered around the center of mass in the $x$ direction \cite{BaiBreen}. The other two, called \textit{dilute boxes}, are placed at a distance $3\ell$ from the center of mass, as shown in Fig.~\ref{fig:ss_MIPS}. A spatial separation of $3\ell$ is essential to avoid the interfacial region, which can introduce large measurement errors \cite{Siebert2018}.
	
	Let $N_+$ and $N_-$ be the number of particles in an $\ell \times \ell$ square region within the dense and dilute boxes, respectively. A suitable order parameter to characterize the phase-separation transition can then be defined as
	\begin{equation}
		\label{eq: MIPS_Order_Par}
		\tp = \frac{1}{\el^2} \avg{\dn_\el} = \avg{\rp}  - \avg{\rn},
	\end{equation}
	where $\avg{\cdot}$ represents the steady-state averages and $\ro_\pm = N_\pm/\el^2$ are the densities of dense and dilute regions as sampled by the selected measurement sub-boxes in a \pss, formally known as the liquid and gas densities. Note that, $\tp$ vanishes in the mixed phase where $\avg{\rp}=\avg{\rn}$.
	
	In this sub-box method, although the simulation runs on an elongated box, the sub-boxes are $\el \times \el$ squares, and the correlation length $\xi$ does not suffer a directional asymmetry at least in a region far from criticality where $\xi \ll 2\el$. However, the fact that the average particle density $\avg{\ro_\el}$ of all four measurement boxes is different from the conserved input density $\ro =\frac{N}{12\el^2}$, may generate an additional loss of precision due to nonconservation. Also, reaching a larger $\el$ to trace the actual behavior of active particle systems is quite difficult. To account for these issues, another approach is considered in Ref.~\cite{Ray2024}, where the aspect ratio was chosen to be 2:1 instead of 3:1, such that the system size is $L_x=2\ell$ and $L_y=\ell$. The order parameter they used, though respects the conserved particle density $\ro$, but the implemented aspect ratio of 2:1 does not rule out additional sampling issues due to the prominent interface overlap. Since an accurate estimate of critical exponents is essential for determining the universality class of a continuous phase transition, we look into the geometric properties of the system from the perspective of percolation theory and try to exploit its connection with the underlying \pst~following the recent work as in Ref.~\cite{Soumya2024}.
	
	Usually, in a \pss~the high-density region is formed by a singly connected component, whereas the low-density region has many disconnected clusters. Here a unique cluster is constructed similar to those defined in the site-percolation problems \cite{Stauffer1979} by connecting two particles if they are nearest neighbors of each other. Although during the evolution of the system the large macro-cluster may break down, a single cluster is always energetically favorable. Thus, a \pst~is always associated with a geometric percolation transition where the size of the largest cluster serves as the order parameter of the system. Note that any configuration of $N$ particles on a lattice can be viewed as a collection of $K$ clusters, indexed as $k=1, 2, \dots, K$, each containing $s_k$ number of particles so that $\sum_{k=1}^K s_k=N$. If $\smx = {\rm max} \{s_k\}$ is the size of the largest cluster, then one can consider $\phi = \smx/L^2$ as the order parameter of the system to characterize the geometric transition, likewise to what is used in site-percolation theory \cite{Stauffer1979, Essam1980}. In the mixed phase, the cluster is typically as large as the finite correlation length $\xi$ (for $\xi\ll L$) of the system, which makes $\phi\to0$ in the thermodynamic limit $L\to\infty$. On the other hand, in the \pss~as the clusters occupy a finite fraction of the particles, thereby becoming comparable to the system size and hence leading to $\phi\ne0$. 
	
	Then the percolation order parameter $\phi$ and susceptibility $\chi$ are defined in a same manner as in ordinary percolation \cite{Stauffer1979}
	\begin{equation}
		\label{eq: Perc_Order_Par}
		\phi = \frac{1}{L^2}\avg{\smx} ;\quad \chi = \frac{1}{L^2}\left( \avg{\smx^2} - \avg{\smx}^2 \right).
	\end{equation}
	The associated critical exponents $\beta$ and $\gamma$ and the correlation exponent $\nu$ [defined in the same way as in Eq.~\eqref{eq: CritExp_Ising}] together determine the universality class of the system. In the context of equilibrium phase transitions, the site-percolation transition of \cg~occurs exactly at the same critical interaction $\Jcs$ where \pst~occurs, but their critical exponents differ \cite{Coniglio1980, Coniglio2001, Fortunato, Janke2005}. The static critical exponents of the \pst~of \cg~in 2D are no different from those of the ferromagnetic transition: $\nuI = 1,~\btI  = 1/8,~\gmI= 7/4$, and hence it can be said that the critical behavior belongs to the \iuc~or $Z_2$  symmetry breaking. Since the percolation transition in \cg~occurs at the same critical point, the correlation length remains the same and hence the associated correlation exponent $\nu$ must remain the same in both transitions. It was argued by Stella and Vanderzande \cite{Stella1989} that even though both the percolation and \pst~share the same critical point, the exponents of the geometric transition cannot be expressed solely in terms of the Ising exponents. They conjectured that the fractal dimension of the percolating cluster at the critical point is not simply $d_f = d - \btI/\nuI= 15/8$; rather it is
	\begin{equation}
		\label{eq: df_relation}
		d_f = d - w \frac{\btI}{\nuI}; \quad  w = \frac{5}{12},
	\end{equation}
	where the additional parameter $w$ is determined from the connection of percolation with tricritical $q=1$-Potts model \cite{Stella1989}. The robustness of this relation is verified in many other models in 2D \cite{Aikya2025}. Thus the exponents of percolation $\{\nu,~\beta,~\gamma\}$ are related to that of the underlying phase transition exponents $\{\nuI,~\btI,~\gmI\}$ as
	\begin{equation}
		\nu =\nuI; \quad \beta= \frac5{12}\btI;\quad \gamma =\frac{13}{12}\gmI,
		\label{eq: perc_Ising_relation}
	\end{equation}
	where $\gamma$ is determined from the scaling relation $2\beta + \gamma =d\nu$ \cite{Stauffer1979, Stauffer1994}.  It is worth mentioning that percolation transition in \cg~with exponents $\nu=1,~\beta=5/96,~\gamma=91/48$, forms a different universality class called interacting percolation or $Z_2$-percolation ($Z_2$P) universality \cite{Coniglio1980,  Coniglio2001, Fortunato, Janke2005}.
	
	Inspired by the observation that percolation properties efficiently capture the nature of the underlying \pst~and that their critical exponents are connected through simple scaling relations, we prefer to study the percolation properties of this IRTP system on a {\it square} lattice and try to infer about the critical exponents of the \pst~using Eq.~\eqref{eq: perc_Ising_relation}. In a few cases, we explicitly study the \m~transition by calculating the order parameter $\tp$ [Eq.~\eqref{eq: MIPS_Order_Par}] from the sub-box densities on a {\it rectangular} lattice ($6\el \times 2\el$). The benefit of studying the percolation transition is that we can consider a square geometry that evades the ill effects of the slab geometry, if any. Moreover, cluster properties can be computed more efficiently in comparison to the determination of high and low densities $\ro_\pm$ and corresponding density fluctuations for the estimation of the critical point and corresponding critical exponents.
	
	\section{Simulations and Results\label{sec: IV}}
	
	In the Monte-Carlo (MC) simulation of the IRTP model, a particle $k$, is chosen at random independently from the collection of $N$ particles; say its position is $\mathbf{i}$ and internal orientation is $\theta_k$ (the unit vector along this direction is  $\bm{\delta}_k = (\cos(\theta_k), \sin(\theta_k))$. The chosen particle can either decide to ``tumble'' with rate $\w$ to change this intrinsic orientation to any of the other three or ``run'' with rate $r={\rm min}\left\{ 1, e^{-\Delta E} \right\}$ to one of its nearest neighbors chosen as follows: site $\mathbf{i} + \bm{\delta}_k$ and other three neighbors in ratio 1:3$p$ distributed equally. Here $\Delta E$ is the change in energy of final and initial configurations, following Eq.~\eqref{eq: E}. The detailed implementation of the dynamics is described in the Supplemental Material \cite{supp}.
	
	\subsection{Percolation of interacting RTPs\label{sec: IVA}} 
	
	We investigate the percolation transition of \rtp~using Monte Carlo simulations of the IRTP model by varying the interaction strength $J$ and the tumbling rate $\w$, while keeping the particle density fixed at $\ro = 1/2$. We also consider a fixed translational diffusion parameter $p = 0.05$, a regime where the re-entrant behavior of the transition is particularly prominent.
	\begin{figure}[t]
		\centering
		\includegraphics[width=\linewidth]{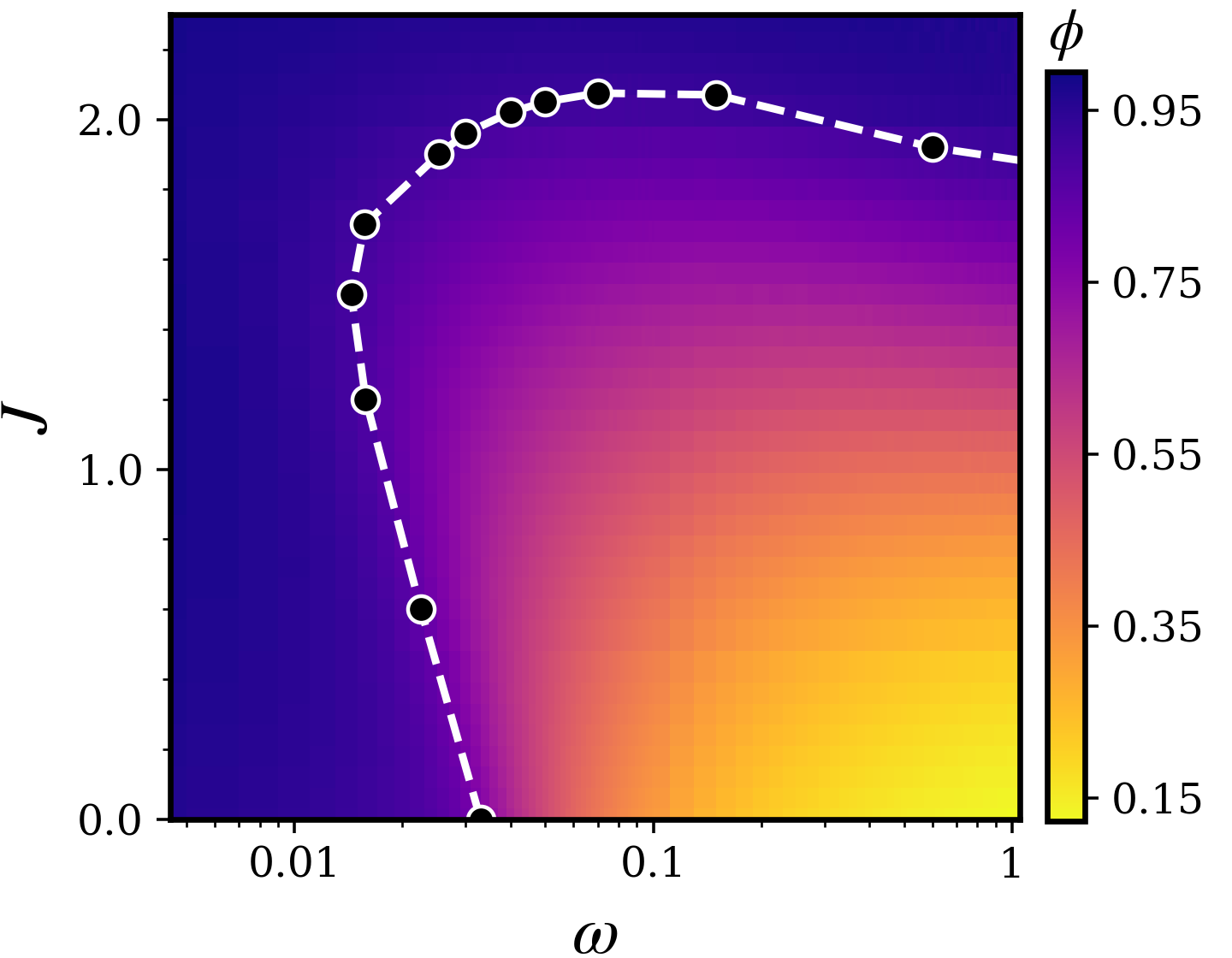}
		\caption{(Color online) Density plot of the percolation order parameter $\phi= \avg{\smx}/L^2$ in the $\w$-$J$ plane. The estimated critical points (circles) are joined by a line; this critical line separates the percolating phase from a nonpercolating one, and it coincides with the critical line of MIPS transition shown later in  Fig.~\ref{fig:phase}. The system size is $64\times64$ and $\ro = 0.5$.}
		\label{fig:Phaseplot}
	\end{figure}
	To study the percolation transition, we choose the order parameter as in Eq.~\eqref{eq: Perc_Order_Par}, generally used to characterize geometric phase separation in percolation problems. A quantity of particular interest is the so-called Binder cumulant $B_L$, defined as
	\begin{equation}
		\label{eq: BC_Perc}
		B_L = 1 - \frac{\avg{\smx^4}}{3\avg{\smx^2}^2},
	\end{equation}
	which is especially useful in the case of continuous phase transitions, because it takes a universal value at the critical point, independent of system size $L$.
	\begin{figure}[t]
		\centering
		\includegraphics[width=\linewidth]{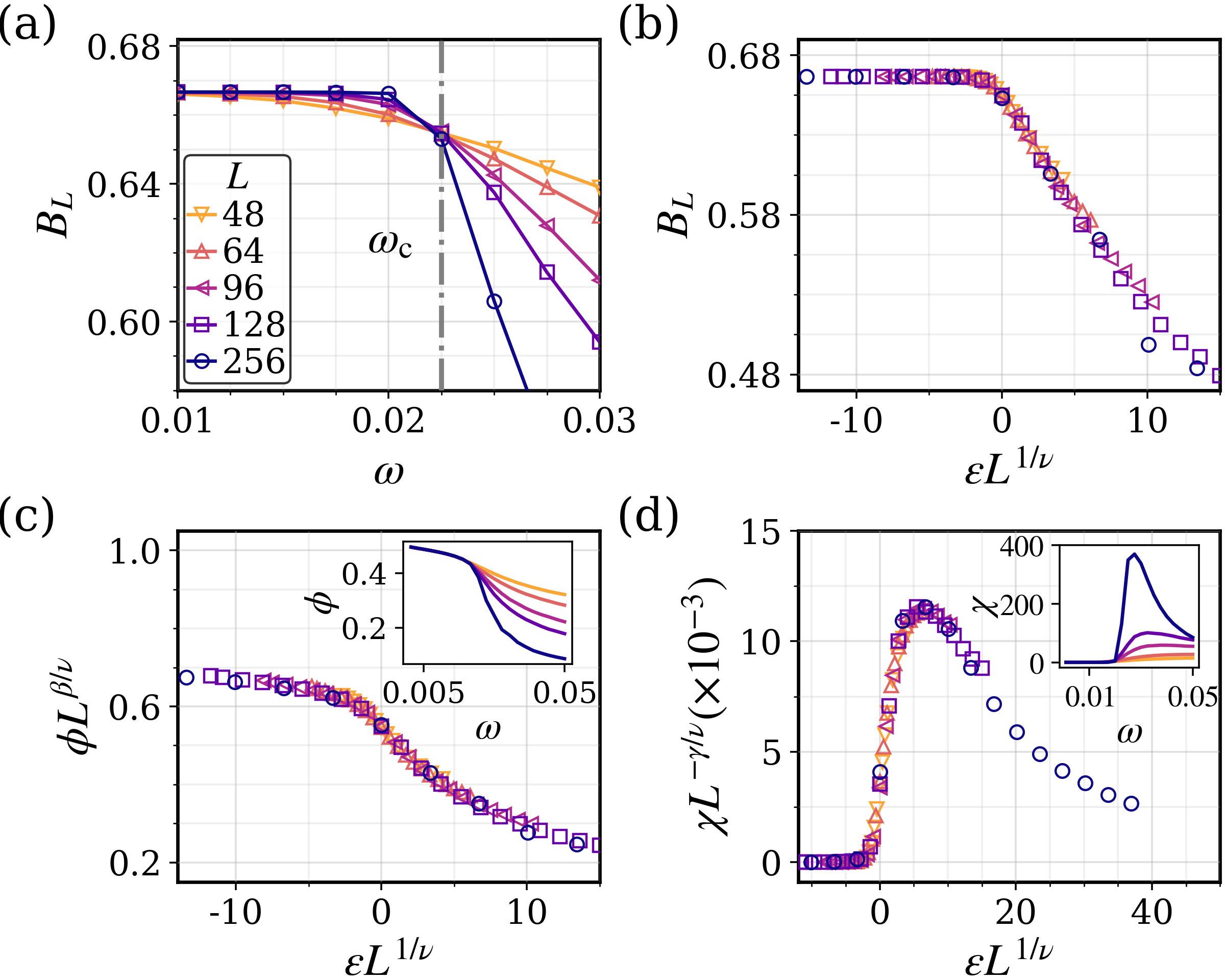}
		\caption{(Color online) Finite-size scaling of percolation transition in IRTP model at $\Jc=0.6,~\rc=\ro=\frac12$ on a   $L\times L$  square lattice.  (a) Estimation of critical point $\wc = 0.0225(2)$ from the crossing point of $B_L$ vs $\w$ curve for different system sizes: $L=48,~64,~96,~128,~256$. Panels (b), (c), and (d) provide scaling collapse of $B_L,~\phi,~\chi$ according to  Eq.~\eqref{eq:FSS}  with  $\nu = 0.82(1)$, $\beta = 0.042(2)$ and $\gamma = 1.556(20).$}
		\label{fig:J = 0.6_RTP_Perc}
	\end{figure}
	For the determination of critical point $\wc$, we then evaluate the Binder cumulant $B_L$ as a function of $\w$ across different system sizes $L$; their intersection point provides a good estimate of the critical value. The critical points calculated from simulations are shown in the $\w$-$J$ plane, in Fig.~\ref{fig:Phaseplot} (symbols connected by a line). The color in
	this figure represents the density of the order parameter $\phi$. For a specific value of $J = 0.6$, the Binder cumulant for different $L$ yields $\wc = 0.0225(2)$, as shown in Fig.~\ref{fig:J = 0.6_RTP_Perc}(a). To calculate the critical exponents of the percolation transition, we employ the finite-size scaling (FSS) relations of $B_L,~\phi,$ and $\chi$,
	\begin{equation}
		\label{eq:FSS}
		\begin{aligned}
			B_L &= f_{B_L}(\varepsilon L^{\frac{1}{\nu}})~; \\
			\phi &= L^{-\frac{\beta}{\nu}}f_{\phi}(\varepsilon L^{\frac{1}{\nu}})~; \\
			\chi &= L^{\frac{\gamma}{\nu}}f_{\chi}(\varepsilon L^{\frac{1}{\nu}})~.
		\end{aligned}
	\end{equation}
	This suggests that $B_L,~\phi L^{\beta/\nu}$ and $\chi L^{-\gamma/\nu}$ must collapse onto a unique scaling function when one plots them against the quantity $\varepsilon L^{1/\nu}$. Since we know $\varepsilon= \w-\wc$, one can use $1/\nu,~\beta/\nu$ and $\gamma/\nu$ as fitting parameters and hence can determine their value as the one which gives the best data collapse.  Following this prescription we obtain the critical exponents $\nu = 0.82(1),~\beta=0.042(2)$ and $\gamma=1.556(20)$, corresponding to $\big(\wc=0.0225,\Jc=0.6\big)$. This is shown in Figs.~\ref{fig:J = 0.6_RTP_Perc}(b), \ref{fig:J = 0.6_RTP_Perc}(c) and \ref{fig:J = 0.6_RTP_Perc}(d) respectively. In the insets of Figs.~\ref{fig:J = 0.6_RTP_Perc}(c) and \ref{fig:J = 0.6_RTP_Perc}(d) we show the raw data.
	
	A closer inspection of the phase plot in Fig.~\ref{fig:Phaseplot} reveals that for small values of $J$ (approximately $J < 1.9$) one can find the critical $\wc$ by varying  $\w$, for a fixed $J$. For larger $J$, this would result in two critical points, as the transition appears to be re-entrant in nature. Then determination of critical parameters would influence each other, resulting in erroneous estimation. To overcome this, we consider changing $J$ for a fixed $\w$; the scaling properties as in Eq.~\eqref{eq:FSS} remain the same, but one must consider $\varepsilon = J-\Jc$. The critical points and the exponents $\nu,\beta,\gamma$ determined from FSS analysis are listed in Table~\ref{tab: RTP_Perc_Table}. Details of the scaling collapse for some of the other critical points are reported in the Supplemental Material \cite{supp}.
	\begin{table}[t]
		\centering
		\setlength{\tabcolsep}{3pt}
		\begin{tabular}{ccccc}
			\hline
			$J_\text{c}$ & $\w_\text{c}$ & $\nu$ & $\beta$ & $\gamma$\\
			\hline
			0.00(0) & 0.0330(2) &0.84(1) & 0.058(3) & 1.564(20) \\
			0.60(0) & 0.0225(2) &0.82(1) & 0.042(2) & 1.556(20) \\
			1.20(0) & 0.0158(2) &0.72(2) & 0.033(1) & 1.374(40) \\
			1.50(0) & 0.0144(4) &0.76(3) & 0.034(1) & 1.452(60) \\
			1.70(0) & 0.0157(2) &0.78(3) & 0.035(2) & 1.490(60) \\
			1.90(0) & 0.0253(2) &0.91(2) & 0.043(2) & 1.734(40) \\
			1.99(2) & 0.0300(0) &0.92(2) & 0.044(1) & 1.752(40) \\
			2.05(2) & 0.0500(0) &0.95(2) & 0.044(2) & 1.812(40) \\
			1.92(2) & 0.6000(0) &1.05(3) & 0.048(2) & 2.044(60) \\
			1.84(2) & 2.0000(0) &1.04(3) & 0.050(3) & 1.980(60) \\
			1.81(2) & 3.0000(0) &1.04(2) & 0.052(3) & 1.976(40) \\
			\hline
		\end{tabular}
		\caption{Critical points and corresponding static exponents of percolation transition of IRTP model at density $\rc = 0.5$.\\
		}
		\label{tab: RTP_Perc_Table}
	\end{table}
	
	The critical exponents can be obtained directly from the following scaling relations. 
	For large $L$, the order parameter scales as $\phi \sim \varepsilon^{\beta}$; thus, a log-log plot of $\phi$ versus $\varepsilon$ yields $\beta$. At the critical point, the finite-size scaling forms are  
	\be
	\phi \sim L^{-\beta/\nu}; \quad 
	\chi \sim L^{\gamma/\nu}; \quad 
	\frac{dB_L}{d\varepsilon}\bigg|_{\varepsilon\simeq0} \sim L^{1/\nu}.
	\label{eq:direct}
	\ee
	Therefore, log-log plots of $\phi$, $\chi$, and $\frac{dB_L}{d\varepsilon}$ as functions of $L$ provide independent estimates of $\beta/\nu$, $\gamma/\nu$, and $1/\nu$, respectively. 
	The results of these fits, for $J=0.6$  are shown in Fig. S11 of the Supplemental Material~\cite{supp}; within error limits, they are in good agreement with the values listed in Table~I.
	
	\subsection{MIPS transition of RTPs\label{sec: IVB}}
\begin{figure}[t]
		\centering
		\includegraphics[width = \linewidth]{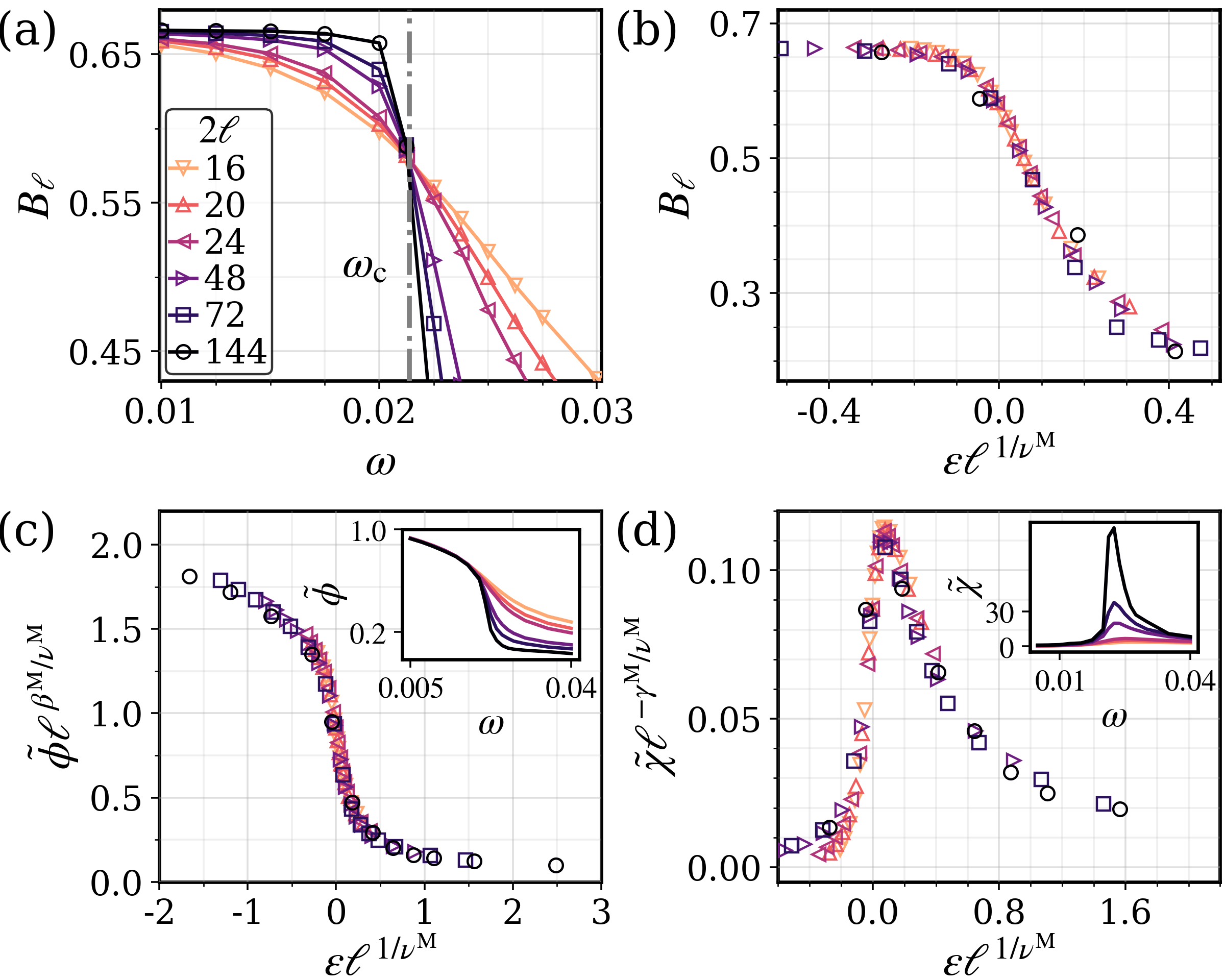}
		\caption{(Color online) Phase separation transition in IRTP model at  $\Jc=0.6$ and, $\ro=\rc=0.5$  on a $6\ell \times 2\ell$ lattice. (a) Estimate of the critical point $\wc=0.0215(2)$ from the crossing point of Binder cumulant $B_{\el}$ (Eq.~\eqref{eq:Binder_Cum_MIPS}) vs $\w$ for different $\el$. Within error limits, $\wc$ matches the same obtained from the percolation transition. Panels (b), (c), and (d) provide estimates of the critical exponents, $\nuM = 0.82(1)$, $\btM=0.12(2)$, $\gmM=1.38(4)$ from finite size scaling similar to Eq.~\eqref{eq:FSS}.}
		\label{fig: J0.6_MIPS}
	\end{figure}

	Following several studies \cite{Fortunato, Stella1989} one can find that the percolation transition in the 2D \cg~model serves as a geometric signature of the underlying magnetic phase transition (\pst). Notably, both the percolation and \pst~share the same critical interaction $\Jcs$ [Eq.~\eqref{eq: Jcs}], and the corresponding critical exponents are linked as outlined in Eq.~\eqref{eq: perc_Ising_relation}. As established in previous studies \cite{Soumya2024} and further confirmed in a later section of this work, the percolation transition in \rtp~belongs to the  $Z_2P$ superuniversality class, while the  \m~transition is believed to fall within the 2D Ising universality class (or $Z_2$ universality) \cite{Partridge2019}.  Since  geometric and \pst~of \rtp~ are  similar to that of the equilibrium  \cg~model, one can draw a correspondence between them, which allows one to reasonably propose that the percolation transition of \rtp~and the underlying  \m~transition should also share the same critical point and an analogous relation between the respective critical exponents:
	\begin{equation}
		\label{eq: MIPS_Perc_Relation}
		\nuM = \nu; \quad \btM = \frac{12}{5}\beta; \quad \gmM = d\nuM-2\btM,
	\end{equation}
	where $d$ is the spatial dimension and the superscript ``M" denotes the exponents belong to \m~transition.
	\begin{figure}[t]
		\centering
		\includegraphics[width=\linewidth]{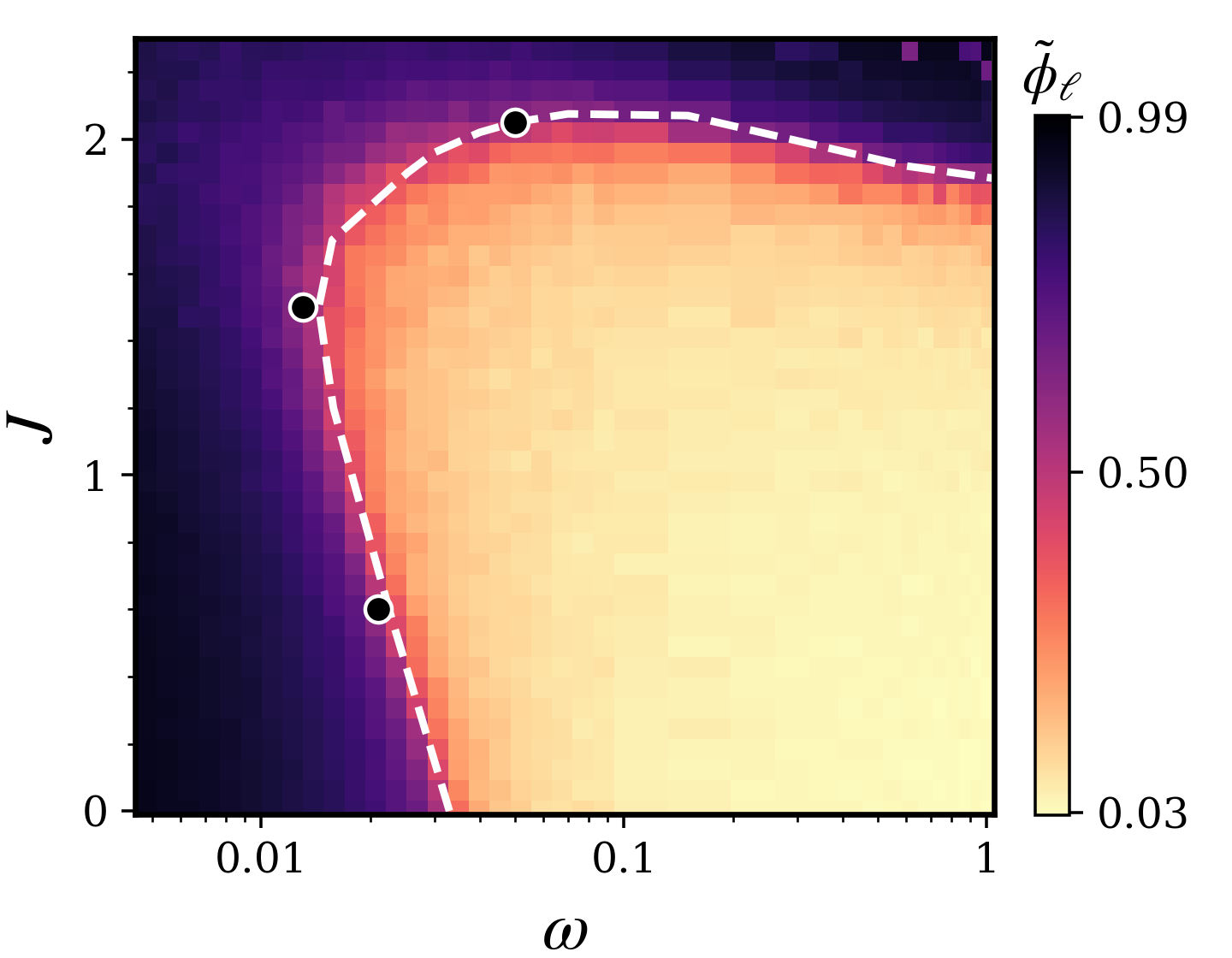}
		\caption{(Color online) Density plot of order parameter $\tp$ [Eq.~\eqref{eq: MIPS_Order_Par}], for a $6\el\times2\el$ system to study \pst~with $\el=36$. Estimated critical points (circles) are shown along with the critical line of percolation transition obtained in Fig.~\ref{fig:Phaseplot}. Within error limits they match well, suggesting that the percolation transition and \pst~occurs simultaneously and they share the same critical points along the critical line. Note that the conserved particle density taken for this case is $\ro=0.5$.}
		\label{fig:phase}
	\end{figure}
	To verify Eq.~\eqref{eq: MIPS_Perc_Relation}, we would need the corresponding static exponents of \m~transition and for that we then employ the sub-box method \cite{Siebert2018} described earlier for the determination of critical point and those static exponents using order parameter $\tp$ as in Eq.~\eqref{eq: MIPS_Order_Par}. For the sake of comparison with the percolation study, here we define the Binder cumulant of $\tp$,
	\begin{equation}
		\label{eq:Binder_Cum_MIPS}
		B_\ell = 1 - \frac{\avg{\dn_\el^4}}{3\avg{\dn_\el^2}^2}.
	\end{equation}
	The critical points estimated from the Binder cumulants
	of MIPS order parameter are shown in Fig.~\ref{fig:phase} (symbols),
	along with the critical line obtained from percolation studies. The color scale in this figure represents the density of
	the order parameter $\tp$. For a representative value $J=0.6$, from the crossing of Binder cumulant $B_\el (\w)$ calculated as
	a function of $\w$ for different sub-box lengths $\el$ (viz. $\el=8,~10,~12,~24,~36,~72$), we find the intersection point to be $\wc=0.0215(2)$ [in Fig.~\ref{fig: J0.6_MIPS}(a)]. This estimate reasonably agrees with the previously obtained value $\wc = 0.0225(2)$ from the percolation study. $B_\el(\w)$ follows a  FSS  relation similar to Eq.~\eqref{eq:FSS}, which is then utilized to obtain the correlation exponent of the \pst~as $\nuM=0.82(1)$ from the scaling collapse [see Fig.~\ref{fig: J0.6_MIPS}(b)].
	
	For the  order parameter  $\tp,$ the susceptibility can be defined as  
	\begin{equation}
		\label{eq:chi_subbox}
		\tilde \chi=\frac{1}{\el^2}\big( \avg{\dn_\el^2}  - \avg{\dn_\el}^2\big).
	\end{equation}
	Both $\tp$ and $\tc$ obey a FSS relation similar to Eq.~\eqref{eq:FSS}; their scaling collapse [described respectively in Figs.~\ref{fig: J0.6_MIPS}(c) and (d)], results in order parameter exponent $\btM = 0.12(2)$ and susceptibility exponent $\gmM = 1.38(4)$.
	
	As an additional check, we apply scaling relations analogous to Eq.~\eqref{eq:direct} to determine $\btM/\nuM$, $\gmM/\nuM$, and $1/\nuM$ for the MIPS transition, as shown in Fig.~S12 of the Supplemental Material~\cite{supp}. The resulting exponents are consistent, within uncertainties, with those obtained from the FSS data collapse: $\{\nuM = 0.82(1),~\btM = 0.12(2),~\gmM = 1.38(4)\}$.
	
	As listed in Table~\ref{tab: RTP_Perc_Table}, the critical exponents obtained from the percolation study of the \rtp~at the critical point $(\wc = 0.0225(2), \Jc = 0.60(0))$, were $\{\nu = 0.82(1),~\beta = 0.042(2), \gamma = 1.556(20)\}$.  Although these two sets of exponents are different, a closer inspection reveals a deeper connection among them. Specifically, the correlation length exponents $\nu$ and $\nuM$ are identical, indicating that the length scales ($\xi$) of the two phenomena are consistent with each other. For the order parameter exponent, while $\beta$ and $\btM$ are not numerically equal, they are related through the scaling relation described in Eq.~\eqref{eq: MIPS_Perc_Relation}. In fact, we observe that $\btM \simeq \frac{12}{5}\beta = 0.11$, which matches with the scaling relation within the margin of error. Additionally, the susceptibility exponents $\gamma$ and $\gmM$ satisfy their respective hyperscaling relations ($\gamma = d\nu-2\beta$ in percolation and $\gmM=d\nuM - 2\btM$ in \m), further supporting the consistency between the two frameworks.
	
	In a similar manner, we computed the critical values: ($\wc=0.012(2),\Jc=1.5$) and ($\wc=0.05,\Jc=2.05(2)$) for the MIPS transition using the sub-box method. These critical points are shown in the $\w$--$J$ plane in Fig.~\ref{fig:phase}, marked as circles. The background shading in the figure represents a density plot of the MIPS order parameter defined in Eq.~\eqref{eq: MIPS_Order_Par}, where darker regions correspond to a more ordered (phase-separated) state.
	For comparison, we also plot the critical line corresponding to the percolation transition, obtained by connecting its critical points determined using the same order parameter [Eq.~\eqref{eq: Perc_Order_Par}]. This line clearly separates the dark (phase-separated) region from the light (mixed) region in the phase diagram. Remarkably, the critical points obtained from the MIPS analysis lie along this same line (within error bars), suggesting that both the percolation and MIPS transitions occur simultaneously upon crossing the critical line. Further, for ($\wc=0.012(2),\Jc=1.5$) and ($\wc=0.05,\Jc=2.05(2)$), we obtain the critical exponents of MIPS phase transition from the sub-box method, which consistently obey Eq.~\eqref{eq: MIPS_Perc_Relation} (for details refer to the Supplementary Material \cite{supp}). These observations indicate that the percolation transition coincides with phase separation, a phenomenon known to be generic in two spatial dimensions \cite{Saberi2010}.
	
	
	\subsection{The phase diagram\label{sec: IVC}}
	
\begin{figure}[t]
		\centering
		\includegraphics[width=\linewidth]{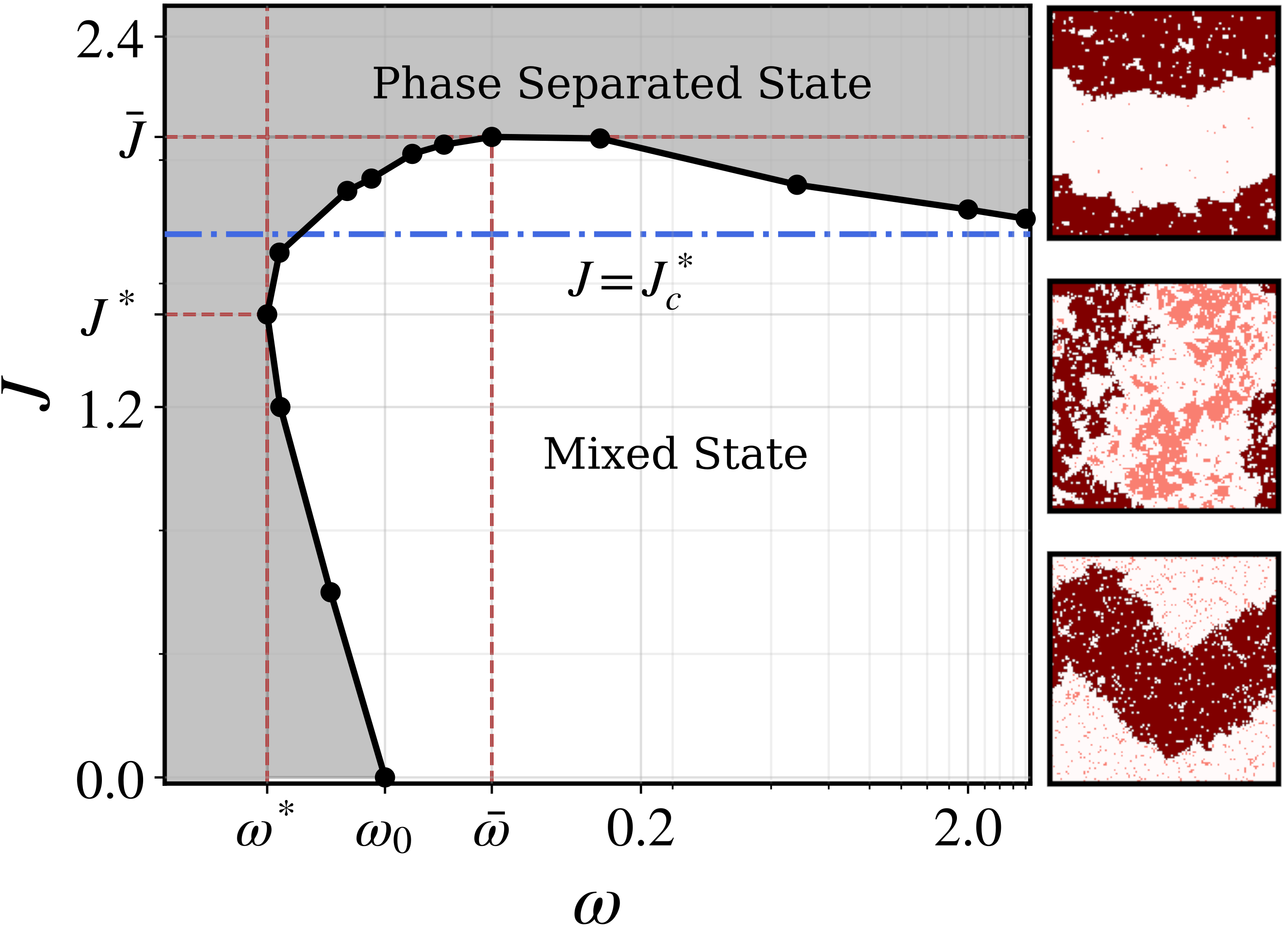} 
		\caption{(Color online) Phase diagram of  the percolation transition in IRTP model in $\w$-$J$ plane for $\ro=0.5$ and $p=0.05$. The critical line is drawn by joining the estimated critical points (dots) from Table~\ref{tab: RTP_Perc_Table}, separates the \pss~from the mixed one. Important features: (i) No transition occurs and the system remains in \pss, when $J > \bar J = 2.075$ or when $\w < \w^* = 0.0144$; note that $J=\bar J$ and $\w =\w^*$ (dashed lines) bound the critical line. (ii) For $J=0$ the system exhibits a \m~transition at $\w = \w_0 = 0.0330(2)$. (iii) In the range $\w^* \le \w \le \bar \w$ the area of the \pss~increases with the increase of motility $\w^{-1}$; phase transitions occurring in this regime may be termed as \m~transition as motility helps the system to phase separate. (iv) For $\w > \bar \w = 0.07$ (at $\w=\bar\w$ the line $J=\bar J$ is tangent to the critical line) the \pss~region starts decreasing with increased motility and the transition here is rather interaction-induced. (v) In the regime $\w^* \le \w \le \w_0$, \m~transition is re-entrant--the \m~phase existing at small $J$ gets destroyed as $J$ is increased and reappears again for higher $J$. The typical configurations of the system in this re-entrant region (at $\w = 0.015$) are shown beside the phase diagram for three different $J=0.0,~1.4,~2.8$ (bottom to top).}
		\label{fig:JW_crit_line}
	\end{figure}

	Building upon the insights gained from the critical exponent analysis, we now turn our attention to the global structure of the phase space governing the percolation transition. Understanding how the transition unfolds across the full range of system parameters provides a comprehensive picture that complements the local scaling behavior discussed in the previous section. The phase diagram shown in Fig.~\ref{fig:Phaseplot} presents a density plot of the order parameter $\phi$ [defined in Eq.~\eqref{eq: Perc_Order_Par}] in the $\w$-$J$ plane. Since the transition occurs within a narrow range of $\w$ (corresponding to high activity/motility), the $\w$ axis is displayed on a logarithmic scale to enhance the visibility of the critical behavior. To further explore the limiting cases of our model, the phase diagram is replotted in Fig.~\ref{fig:JW_crit_line}. The horizontal dashed-dotted line indicates $J = \Jcs = 2\ln(1 + \sqrt{2})$, the critical interaction strength for equilibrium phase separation in the corresponding \cg~model. As $\w \to \infty$, the system becomes passive, and accordingly, the critical line approaches this equilibrium limit: $\Jc \to \Jcs$.
	
	Several limiting cases offer deeper insights. When $p = 1$, particles move isotropically, with their internal orientation playing no role in the dynamics. In this case, the model reduces to the simple case of 2D \cg~model, which undergoes phase separation at $\Jcs$. Thus, this horizontal line also represents the critical line of the IRTP model in the $p = 1$ limit. In contrast, setting $p = 0$ reduces our model to the one studied in Ref. \cite{Ray2024}, where \pst~occurs when $J$ crosses a threshold $\Jc(\w)$ which is larger than $\Jcs$ for any finite $\w$; thus a stronger attractive interaction is required for the particles to phase separate. The \pst~observed in the absence of any translational diffusion ($p = 0$) may be termed as \textit{interaction-induced phase separation}, as increased activity $(\w^{-1})$ here suppresses cluster formation and the system needs interaction to phase separate. This suppression is likely due to an effective repulsion induced by the active motion of attractive particles, as suggested in Ref. \cite{Urna2025}.
	
	However, when translational diffusion is introduced ($p > 0$), the nature of the critical line changes. For example, with $p = 0.05$ \m~transition occurs even at weak or zero attraction, as shown in Fig.~\ref{fig:JW_crit_line}. The shaded region in the figure represents the \pss, bounded by a critical line (solid line with circles). For the noninteracting case ($J = 0$), now a \m~transition occurs at $\w = \w_0 = 0.0330$, in line with the findings of Ref. \cite{Soumya2024}, which demonstrated that \m~can arise at finite tumble rates when $p > 0$.
	
	The horizontal dashed line $J = \bar J$ and the vertical dashed line $\w = \w^*$ are tangent to the critical line at $(\bar \w =0.07, \bar J=2.075)$ and $(\w^*=0.0144, J^*=1.5)$ respectively, which means the IRTP system phase separates for any $J > \bar J$ (irrespective of $\w$) and any $\w < \w^*$ (irrespective of $J$). In the regime $\w^* < \w < \bar \w$ -- motility does enhance the stability of the \pss, as the region of \pss~increases with increased motility. Due to this fact, the \pst~occurring here can be termed as \m, and on the other hand motility suppresses cluster formation when $\w > \bar \w$; here the \pst~is interaction induced.
	
	Another interesting feature that can be observed from the phase diagram is that we have a re-entrant phase transition in the regions (a) $\w^* \le \w \le \w_0$ and (b) $\Jcs \le J \le \bar J$. In the first case, when one keeps increasing $J$ while keeping $\w$ fixed, the system first enters from a \pss~to a mixed one and transits again to the \pss~with a further increase of $J$. The same happens in the second case when $\w$ is increased keeping $J$ fixed. Although the \pst~of the IRTP model is studied at a specific value of $p$, which is $0.05$, it is to be noted that this choice is in no way special; we expect the qualitative features of the phase diagram to persist for any  $p > 0$.
	
	\section{Role of density\label{sec: V}}
	
	So far we have studied the percolation and \m~transition of \rtp~at conserved particle density $\ro=1/2$, expecting that the critical density is not very different from it; in fact, $\ro=1/2$ is indeed the exact critical density for the passive counterpart (2D \cg), which obey $\tp \to - \tp$  (or particle-hole) symmetry. Due to the absence of such symmetry in the IRTP model, to estimate the critical point and the critical exponents, one needs to tune two parameters ($\ro$ and $\w$) simultaneously for a fixed interaction strength $J$. For such a situation, a finite-size scaling method was first proposed for Ising-like models in Ref. \cite{Binder1981} and later put forward to liquid-gas phase transitions in Ref. \cite{Rovere1990}. The basic idea was to divide the system of size $L$, into square sub-boxes of size $\el=(L/n)^d$, where $n$ is an integer and $d$ is the linear dimension. Now, even though the overall conserved density $\ro$ is fixed, these sub-boxes possess their own fluctuating number of particles, leading to a steady-state subsystem density distribution function $P(\ro)$. 
	\begin{figure}[t]
		\centering
		\includegraphics[width=\linewidth]{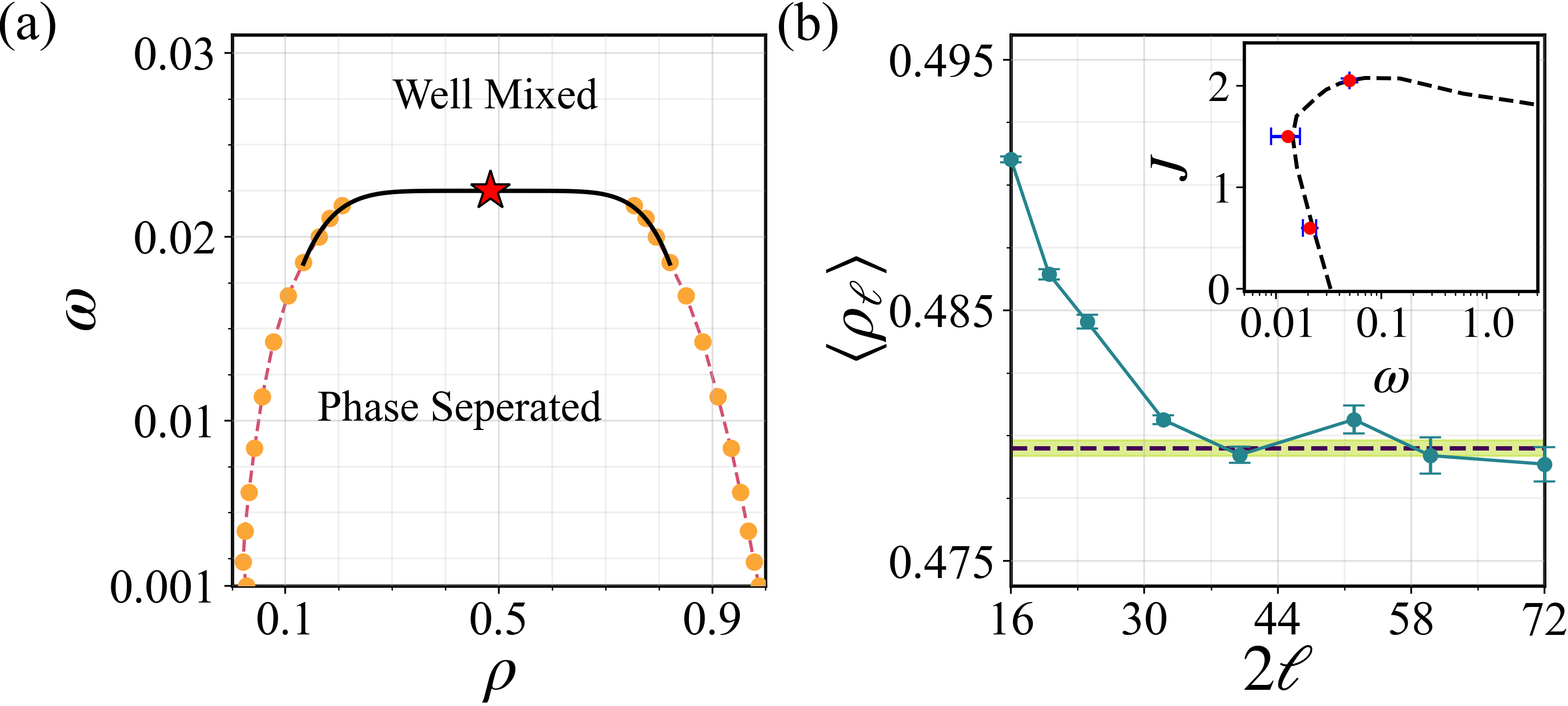}
		\caption{(Color online) Coexistence line and critical density estimation for IRTP model for $J=0.6, p=0.05$: (a) Coexistence densities $\rp$ (right branch) and $\rn$ (left branch) for different $\w,$ obtained  from  $6\ell\times 2\ell$ system using sub-box method are shown as filled circles; the error bars are smaller than  the  point size. The critical point ($\rc=0.4795(3)$, $\wc=0.0225$) is denoted as star. Points far from the critical point are connected through a dashed line as a guide to the eye and for the close ones, both the liquid and gas branches are fitted with a power law, where the exponent $\btM = 0.15$ matches well within the error bounds with the one obtained from the FSS collapse of $\tp$ [Eq.~\eqref{eq: MIPS_Order_Par}].  (b) Average density $\avg{\ro_\el}$ of sub-boxes is plotted with sub-box size $\el.$  The dashed line is the estimate of $\rc=0.4795(3)$ and the shaded region is the error in it. In the inset, critical points [viz. ($\wc=0.021,\Jc=0.6$),($\wc=0.013,\Jc=1.5$),($\wc=0.05,\Jc=2.05$)] obtained with newly estimated densities are shown with their respective uncertainties. The dashed line is constructed with the critical points obtained with $\rc =\frac12$ from Table~\ref{tab: RTP_Perc_Table}.}
		\label{fig: J0.6_Dens_Var}
	\end{figure}
	
	This approximation of $P(\ro)$ in Refs.~\cite{Binder1981, Rovere1990}, however, suffers from interface overlap between two phases. In the subsequent improved work \cite{Siebert2018}, an elongated simulation box of aspect ratio 3:1 was taken as described in Sec. \ref{sec: III}. With this setup, in the \pss~the bulk liquid and gas phases can be accurately sampled through the four sub-boxes while avoiding considerable overlap with the in-between interface region comprising parts of the liquid and gas phases. In this elongated geometry, the edges of the bulk phases (each of an area $\el \times 2\el$) are $2\el$ apart from one another, and each edge lies on an average $\el$ distance away from the interface. Due to the isotropic growth (as expected for systems belonging to \iuc), the correlation length $\xi$ is forced to grow up to a maximum of $2\el$, which is the shorter length of the system. So in the region $\w<\wc$, when $\xi \ll 2\el$, the interfacial fluctuation would not distort the bulk phases and thus the samples collected from liquid (dense sub-box) and gas (dilute sub-box) become uncorrelated.  Then  $P(\ro)$  constructed from the independent samples collected from all four sub-boxes would result in a double-peaked profile.
	
	Contrary to this in the region $\w>\wc$, density fluctuations are of the order of $\xi$, which itself is very small ($\xi \ll 2\el$), thereby making the density samples of the four boxes homogeneous. 
	Corresponding $P(\ro)$ is expected to be a Gaussian centered at the system's average density $\avg{\ro_\el}$. Further, in the critical region when the correlation length $\xi$ grows to an extent that is comparable to the sub-box size $\el$ ($\xi \simeq \el$), and due to that the distribution becomes non-Gaussian but follows a universal scaling function \cite{Rovere1990} 
	\begin{equation}
		\label{eq:Den_Scaling}
		\begin{split}
			P(\ro) &= \el^{\btM/\nuM} \, {\cal F}\big( \Delta\ell^{\btM/\nuM}, \el^{1/\nuM}|\varepsilon| \big),
		\end{split}
	\end{equation}
	where ${\cal F}(.)$ is a scaling function, $\Delta=(\ro - \rc)$, and $\varepsilon=\w-\wc$.
	\begin{figure}[t]
		\centering
		\includegraphics[width=\linewidth]{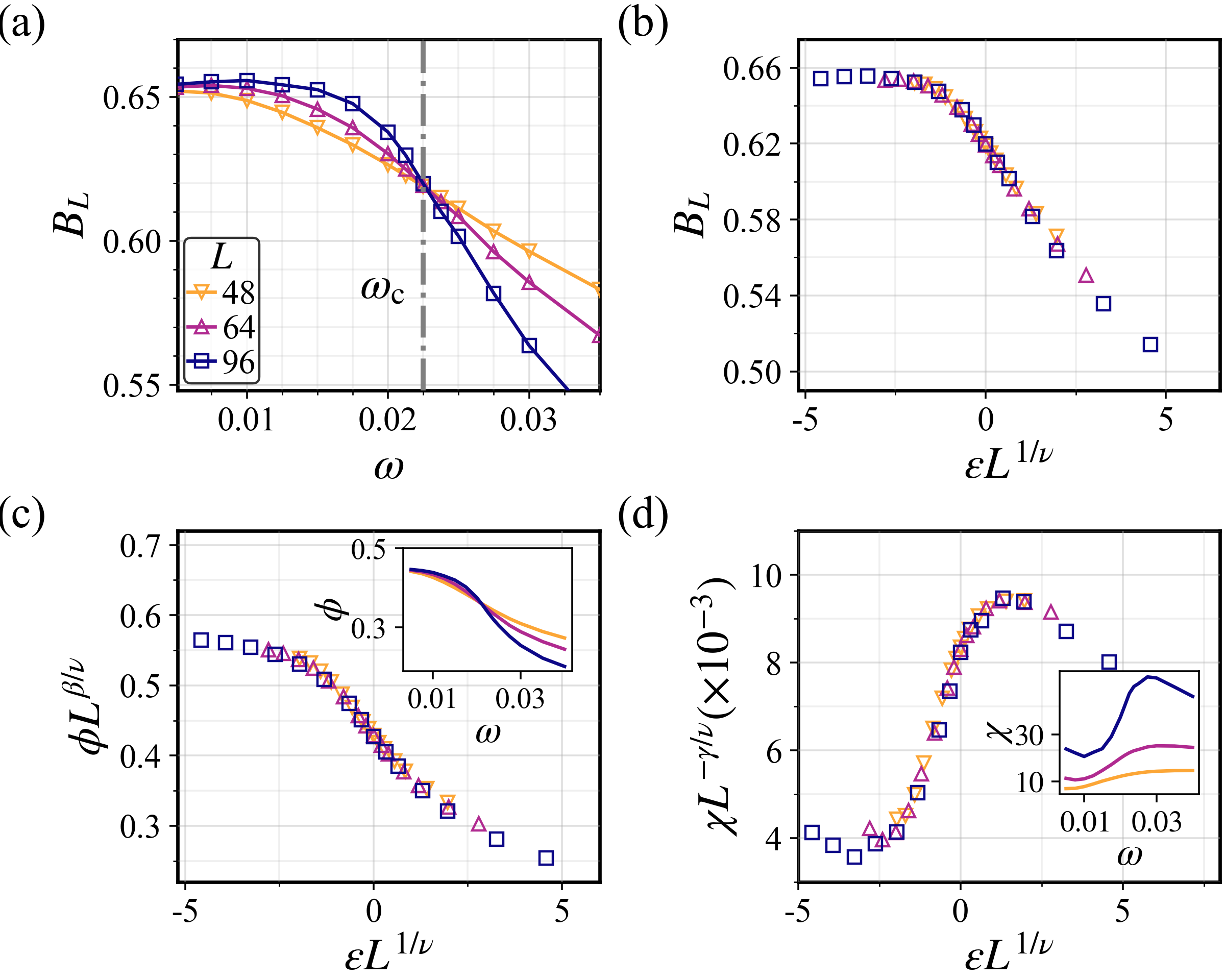}
		\caption{(Color online) FSS collapse for percolation transition of IRTP model on a $L \times L$ square lattice, with interaction strength $\Jc=0.6$ at estimated critical density $\rc=\avg{\ro_\el}=0.4795(3)$. The critical point shown by the dashed line in (a) is the same as obtained in the case of density $\rc=0.5$ in Fig.~\ref{fig:J = 0.6_RTP_Perc}(a), which is $\wc = 0.0225(2)$ for this case. Similarly from (b), (c) and (d) we estimated the critical exponents $\nu,~\beta,$ and $\gamma$ to be the same as that of Fig.~\ref{fig:J = 0.6_RTP_Perc}, as in this case, they are $0.82(2),~0.042(2)$ and $1.556(40)$ respectively.}
		\label{fig:FSS_J0.6_r0.4795}
	\end{figure}
	This universal scaling form asserts that choosing $\ro=1/2$ as the critical value does not compromise the accurate estimation of the other critical parameters and exponents. This assertion is well supported by the convergence of Binder cumulant ($B_\el$ or $B_L$) for both the sub-box method and percolation study at a single point [see Fig.~\ref{fig:J = 0.6_RTP_Perc}(a) and Fig.~\ref{fig: J0.6_MIPS}(a)], otherwise the density-dependent corrections in Eq.~\eqref{eq:Den_Scaling} would have prevented it.
	
	Now for the determination of the actual critical density, we note that the average density $\avg{\ro_\el}$ of all the four sub-boxes measured at the critical point $\wc$ saturates to a critical value $\rc$ as $\el \to \infty$. For $\Jc=0.6$, we see from Figs.~\ref{fig: J0.6_Dens_Var}(a) and \ref{fig: J0.6_Dens_Var}(b) that the average density $\avg{\ro_\el}$ corresponding to different $\el$, calculated by averaging density samples obtained from all independent runs and over all sub-box sizes, saturates to $\rc=0.4795(3)$.  To ascertain that $\wc$ and the critical exponents remain the same within the error limits, we again study the percolation transition of the IRTP model at $\Jc=0.6$, with particle density set at this newly estimated critical value $\rc = 0.4795$. From the crossing point of $B_L$ vs $\w$ curves in Fig.~\ref{fig:FSS_J0.6_r0.4795}(a) we get $\wc=0.0225(2)$ which matches well with the $\wc$ value listed in Table~\ref{tab: RTP_Perc_Table} for $\ro=1/2$. Moreover, $B_L,~\phi L^{\beta/\nu}$ and $\chi L^{-\gamma/\nu}$ exhibit good data collapse when plotted against $\varepsilon L^{1/\nu}$, for $\nu=0.82(2),~\beta=0.041(3)$ and $\gamma=1.558(40)$ [see Figs. \ref{fig:FSS_J0.6_r0.4795}(b), \ref{fig:FSS_J0.6_r0.4795}(c), and \ref{fig:FSS_J0.6_r0.4795}(d) respectively] which agree well with the critical exponents listed in Table~\ref{tab: RTP_Perc_Table} within their respective margins of error.
	
	These findings confirm that the estimated critical parameters at $\rc = 0.4795(3)$ are consistent with those obtained at $\rc = 0.5$. The agreement across different system sizes and methods demonstrates that the finite-size scaling analysis remains valid even with this refined estimate and thereby matches well with the assertion given in Refs.~\cite{Rovere1993, Partridge2019, Soumya2024}.
	
	\subsection{The coexistence line and the critical density\label{sec: VA}}
	
	The density samples of the dense and dilute boxes $\ro_\pm$ estimated earlier can be utilized to obtain the phase-coexistence line of the system. For $\Jc=0.6$, we have shown the plot of the coexistence line for different $\w$ values in Fig.~\ref{fig: J0.6_Dens_Var}(a). As $\w$ approaches its critical value $\wc=0.0215(2)$, both $\rp$ and $\rn$ are expected to reach a unique value $\rc$ but the correlation length $\xi$ of the system also becomes larger at the critical point. For that reason, the high and low-density regions are no longer well separated due to substantial overlap from the interface region at the boundaries, and it becomes increasingly difficult to distinguish $\rp$ from $\rn$. We try to find the best-fit line that connects passes through $\ro_\pm$ points determined for $\w < \wc$. The curvature of the best-fit line (represented by a solid black line) near its maxima at $\rc=0.4795$, determined in the previous section, is quite small, as expected. Indeed, $\rc$ is not much different from $\ro=0.5$; also, the critical exponents calculated at $\rc=0.4795$ do not differ appreciably from the same obtained at $\ro=0.5$.
	\begin{figure}[t]
		\centering
		\includegraphics[width=0.9\linewidth]{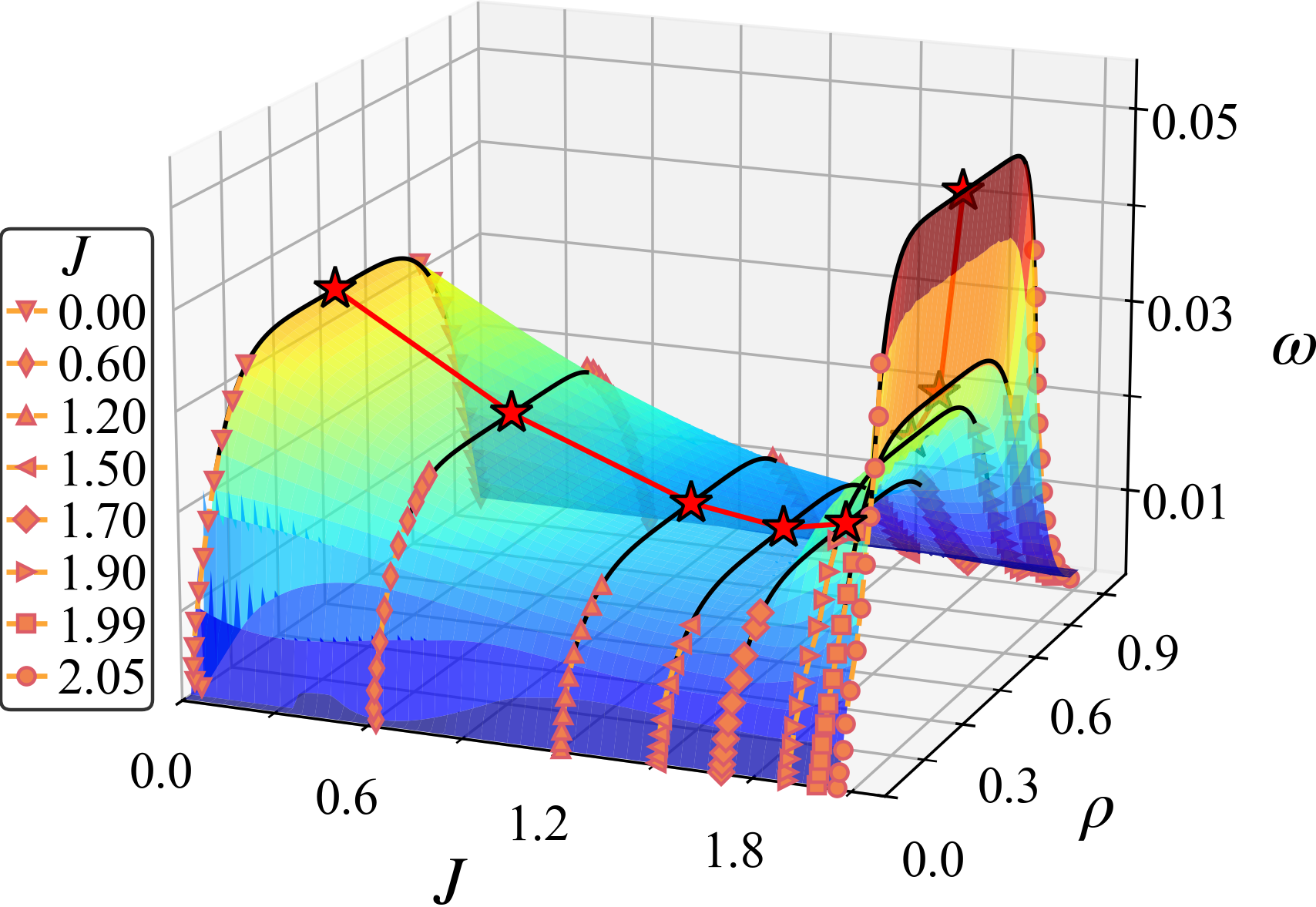}
		\caption{(Color online) The $3$D phase diagram of IRTP model in $J$-$\ro$-$\w$ phase space. Interpolation of the coexistence lines in the $\ro$-$\w$ plane for different $J$ values leads to a surface that separates the mixed phase from the \pss. The zero-gradient condition forms a line (the critical line) on this surface. The critical points $(\Jc,\rc,\wc)$ obtained from simulations align well with this critical line.}
		\label{fig:3d}
	\end{figure}
	
	The coexistence lines for different $J$ values are shown together in Fig.~\ref{fig:3d} in the 3D phase space $J$-$\ro$-$\w$, with the critical points marked as stars. An interpolated surface passing through these coexistence lines separates the \pss~(below this surface) from the well-mixed (above the surface) state. The critical line, obtained by joining these critical points (stars), is also shown in the 3D phase space. Initially, the area of the \pss~decreases with increasing $J$, which is counterintuitive, as the \m~state existing for $J=0$ (the case of noninteracting \rtp) is destroyed as one introduces attraction among the particles. These results are, however, consistent with earlier works \cite{Redner2013, Redner2013a, Soto2014, Whitelam2018, OByrne2022} of a system of active colloids with attractive interaction. It is also consistent with the exact results of Ref.~\cite{Urna2025} obtained for two active particles, where the authors show that attractive interaction generates effective repulsion among the particles, and works against large cluster formation. For very large $J$, the effect of activity is insignificant and at that limit, phase separation occurs rather due to attractive interaction, not due to motility; hence it can be concluded that \pst~for large $J$ is only interaction induced, similarly to what one sees in equilibrium lattice gas models.
	
	\section{Superuniversality of Percolation and MIPS transitions\label{sec: VI}}
	
\begin{figure}[!htbp]
		\centering
		\includegraphics[width=\linewidth]{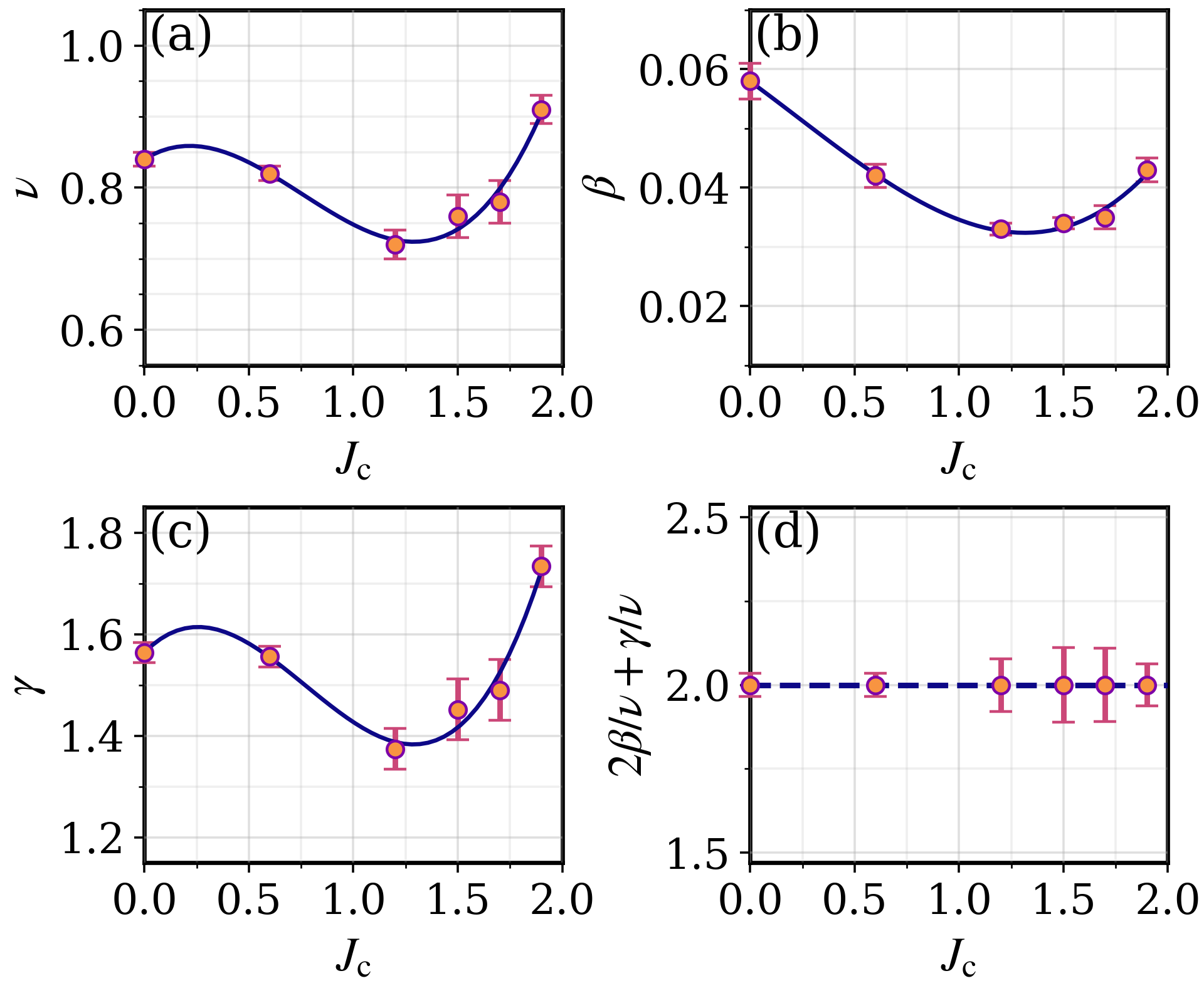}
		\caption{(Color online) Continuous variation of the percolation exponents: (a) $\nu$, (b) $\beta$ and (c) $\gamma$ of IRTP model along the critical line shown as a function of $\Jc$ (taken from Table~\ref{tab: RTP_Perc_Table}). The solid lines are cubic fit to the existing data. Panel (d) shows that the exponents obey the hyperscaling relation $d = 2\beta/\nu + \gamma/\nu$.}
		\label{fig:nu_bt_gm_df}
	\end{figure}

	So far, we have studied the percolation transition of interacting RTPs on a square lattice and established its connection to the associated MIPS transition via Eq.~\eqref{eq: MIPS_Perc_Relation}. The critical behavior of the percolation transition of the IRTP model, summarized in Table~\ref{tab: RTP_Perc_Table}, reveals continuously varying critical exponents: $\nu$, $\beta$, and $\gamma$ vary within the ranges $(0.72\,\text{--}\,1.05)$, $(0.033\,\text{--}\,0.058)$, and $(1.374\,\text{--}\,2.044)$, respectively. Although some of these exponents change by as much as $30\%$, they consistently satisfy the hyperscaling relation $2\beta + \gamma = d\nu$ within the margin of error (see Fig.~\ref{fig:nu_bt_gm_df}).
	
	As expected--and verified explicitly for four representative cases ($\Jc = 0.0, 0.6, 1.5$, and $2.05$)--the corresponding MIPS transition is governed by the same set of critical exponents, which are related to those of percolation via Eq.~\eqref{eq: MIPS_Perc_Relation}. Consequently, the critical exponents associated with the MIPS transition also exhibit continuous variation, as reported in Table~\ref{tab: exponents_connection}. This table further includes a comparison with critical exponents previously reported for MIPS transitions in other active-matter models in two dimensions. Notably, while the values vary across models, they fall within the same range of continuously varying exponents observed in our study.
	
	\begin{table}
		\centering
		\setlength{\tabcolsep}{3pt}
		\begin{tabular}{lccc}
			\hline
			Model & $\nuM$ & $\btM$ & $\gmM$ \\
			\hline
			CLG  & 1 & $1/8$ & $7/4$ \\
			RTPs (Model-I) \cite{Dittrich2021} & 0.98 & 0.15 & 1.66 \\
			RTPs (Model-II) \cite{Dittrich2021} & 0.98 & 0.22 & 1.68 \\
			RTPs (triangular) \cite{ Partridge2019} & 1 & -- & 7/4 \\
			ABPs (continuum) \cite{Maggi2021} & 1.03(10)  & 0.133(22) &  1.84(20)\\
			IRTPs (this work) & 0.72--1.05& 0.079--0.139 & 1.282--1.822 \\
			\hline
			
		\end{tabular}
		\caption{Comparison of critical exponents of MIPS transition  in 2D.}
		\label{tab: exponents_connection}
	\end{table}

	\begin{figure}[t]
		\centering
		\includegraphics[width=\linewidth]{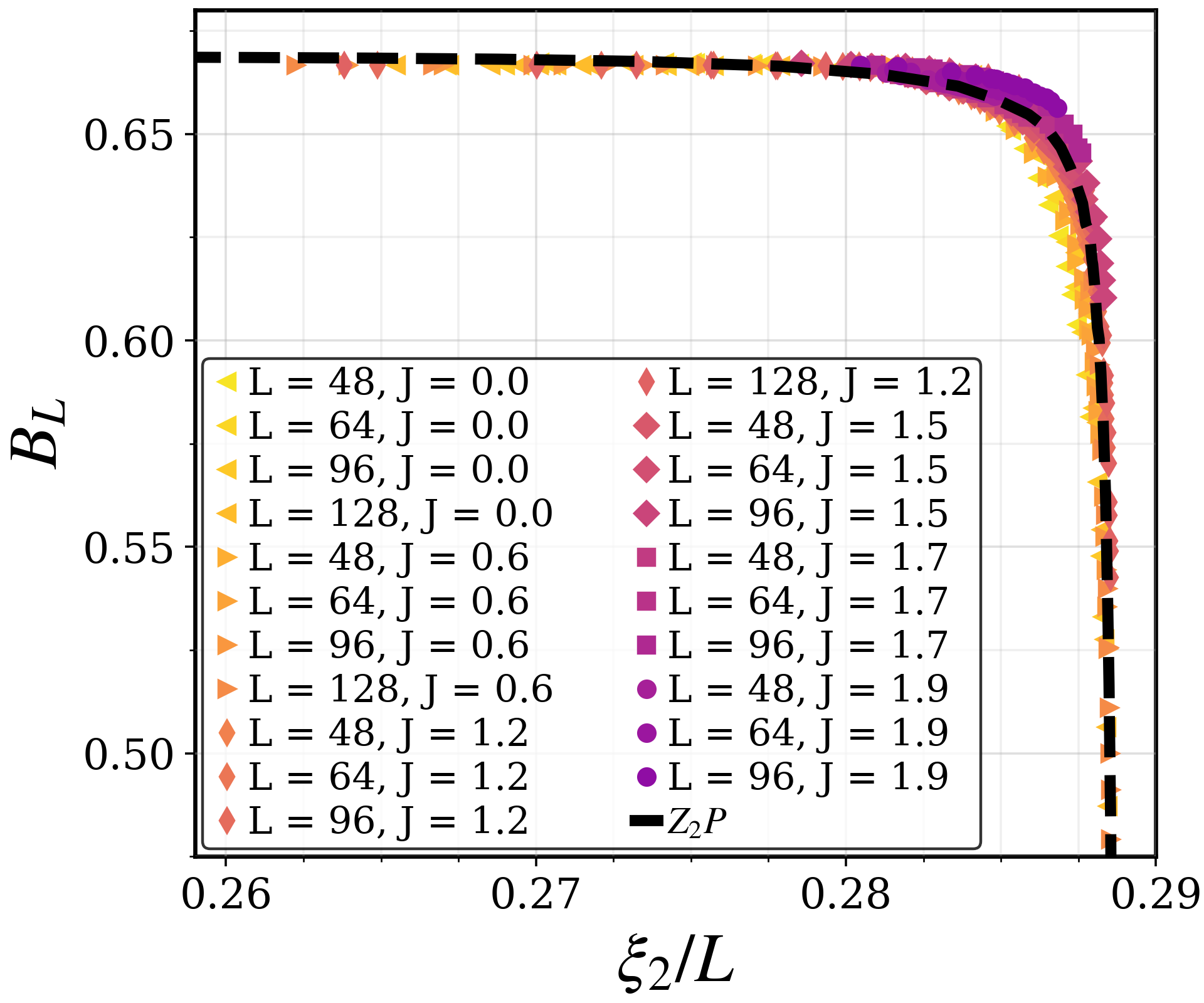}
		\caption{(Color online) Plot of $B_L$ vs $\xi_2/L$ as a function of $\w$ for different values of $J$ and $L$. The dashed solid line represents the same scaling function of $Z_2P$ universality class, obtained by simulating 2D lattice gas model.}
		\label{fig:xi2_col}
	\end{figure}
	
	These continuously varying critical exponents suggest the existence of a marginal operator in the system (identifying such a marginal operator is beyond the scope of this work though). Recently it was proposed that systems with continuously varying critical exponents form a superuniversality class in the sense that certain scaling functions remain invariant all along the critical line; these scaling functions are identical to that of the parent universality class \cite{Vicari, Mukherjee2023}. One such scaling function is the Binder cumulant $B_L$ expressed as a function of a variable $\xi_2/L$, where $\xi_2$ represents the second moment of the correlation length $\xi$, defined as follows  
	\begin{equation}
		(\xi_2)^2 = \frac{\int_0^\infty r^2C({\bf r})\,d{\bf r}}{\int_0^\infty C({\bf r})\,d{\bf r}};
		\quad C({\bf r})  = \langle  n_{\bf i}  n_{\bf i + r} \rangle  -\ro^2. \nonumber
	\end{equation}
	As Binder cumulants are already calculated along the neighborhood of the critical line to estimate the critical thresholds ($\wc,\Jc$), it only remains for us to calculate $\xi_2$ from the pair correlation function $C(r)$ (where $r= |{\bf r}|$), obtained through MC simulations near the critical line. In Fig.~\ref{fig:xi2_col}, a plot of the Binder cumulants $B_L$ corresponding to different interaction strengths against the quantity $\xi_2/L$ is shown as a parametric function of different critical points $(\wc, \Jc)$ for different $L$ values. It is evident from the figure that these curves match quite well with each other irrespective of the value of the critical threshold and the system size; more importantly, independent of the value of the critical exponents which vary continuously along the line. Along with these superuniversal curves, we have also plotted the corresponding scaling function of $Z_2P,$ obtained here for correlated site percolation in 2D \cg~model at criticality (shown by the dashed line in Fig.~\ref{fig:xi2_col}). A very good alignment among the scaling functions of the IRTP model with that of the  CLG model provides clear evidence that the geometric transition of \rtp~in 2D forms a  $Z_2P$ superuniversality class.

	Further, since the critical exponents $\nuM,~\btM,~\gmM$ of the underlying \m~transition are only scaled variants of the percolation critical exponents and follow  Eq.~\eqref{eq: MIPS_Perc_Relation}, they also vary continuously along the critical line in a similar fashion. Consequently, we conclude that the \pst~in IRTP model belongs to $Z_2$ superuniversality class. (or Ising superuniversality class).
	
	\section{Discussion\label{sec: VII}}
	\begin{figure*}
		\centering
		\includegraphics[width=0.95\linewidth]{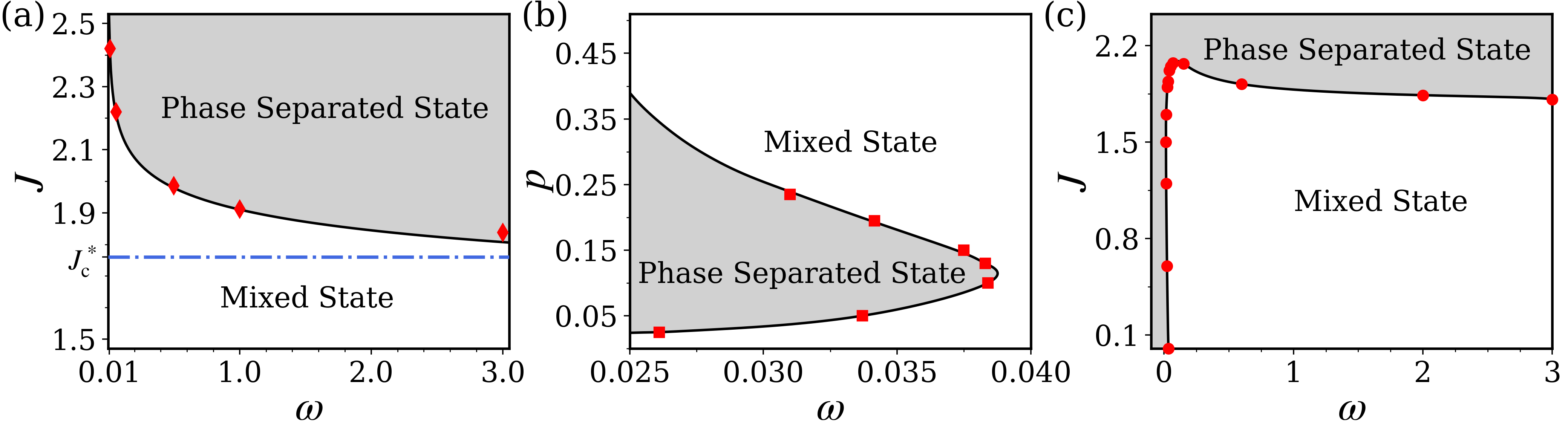} 
		\caption{(Color online) Phase diagram of IRTP model for $\ro=0.5$ in (a) $\w$-$J$ plane (for $p=0)$, (b) $\w$-$p$ plane (for $J=0$), and (c) $\w$-$J$ plane (for $p=0.05)$. In all three cases, estimated critical points (symbols) are joined by best-fit lines (the critical line) that separate the \pss~from the mixed ones.}     \label{fig:all_phase}  
	\end{figure*}
	In our IRTP model, although we choose the Metropolis rate following the energy function (Eq.~\eqref{eq: E}) identical to the equilibrium \cg~models, the system does not obey detailed balance due to the directional asymmetry in their motion for any  $0\le p<1$. This asymmetry forces the system to be in a nonequilibrium steady state, and the phase separation for any $p<1$ occurs as a nonequilibrium phase transition. Also, depending on all the parameters of the model: the conserved particle density $\ro$, tumbling rate $\w$ (or activity/motility $\w^{-1}$), translational diffusion $p$, and attractive interaction strength $J$, our extensive numerical simulations result in the following conclusions.
	
	(i) In the absence of any translational diffusion ($p=0$), motility hinders cluster formation, as a larger attractive interaction is required with increased motility, i.e., $\Jc(\w) >  \Jc(\infty);\quad\forall~\w>0$. A representative figure is shown in Fig.~\ref{fig:all_phase}(a). In this case, we only have an interaction-induced phase separation transition.
	
	(ii) For $J=0$, we observed a \pst~in the $\w$-$p$ plane which is purely motility induced. Since the system is well mixed for all $\w$ when $p=0$ (as discussed above) and $p=1$ (the model reduces to the symmetric exclusion process where all configurations are equally likely), the \m~transition must be re-entrant: For small $\w~(>0)$ the system starts with a mixed phase at $p=0$ and enters a \pss~as $p$ is increased, further increase of $p$ destroy the \pss~leading to another phase transition (\pss~to mixed state). Figure~\ref{fig:all_phase}(b) demonstrates this $J=0$ situation. 
	
	(iii) When $p \ne 0$ but fixed, there is an \m~transition at $J=0$, where \rtp~phase separate when $\w$ decreases below a threshold $\wc>0$. Although there is no explicit interaction provided externally, particles generate an effective nonreciprocal interaction because of their high motility (low $\w$), which is responsible for the transition. We find that any additional attraction $J>0$ acts against cluster formation: The \pss~existing at $J=0$ is destroyed when $J$ is increased keeping $\w$ fixed--a further increase of $J$ results in a \pst~again; this time it is interaction induced, however. The phase diagram is described in Fig.~\ref{fig:all_phase}(c) for $p=0.05$.

	In a phase separation transition, the system exhibits coexistence between two distinct densities: $\rp$ (dense or liquid phase) and $\rn$ (dilute or gas phase), with $\rp > \rn$. These densities vary continuously along the coexistence line as external parameters such as $\w$ is tuned.   At certain  $\wc$ where the two densities merge, i.e., $\ro_\pm = \rc$,is   a critical point;  for $\w>\wc$ the systems maintains a homogeneous density profile. Since $\ro_\pm$ form separate branches, their merging at the critical point implies a vanishing slope at $\rc$. In absence of  particle-hole symmetry, like in the IRTP model,  the critical density need not be $\rc = 1/2;$ however,  the deviation is small for the parameter values explored in our simulations. Furthermore, we have verified that the critical exponents we estimate remain robust and are insensitive to the precise value of $\rc$, provided it lies close to $1/2$. This insensitivity may be a consequence of the shallow curvature of the coexistence line near its extremum. A complete phase diagram of the model in the $J$–$\ro$–$\w$ phase space is presented in Fig.~\ref{fig:3d} for $p = 0.05$, where a domelike coexistence surface separates the phase-separated region from the mixed (homogeneous) phase. The maximal points on this surface define a critical line.

	We also investigate the nature of criticality along the critical line to determine the universality class of the \pst~in IRTP model. It turns out that the critical exponents vary continuously along the critical line, similar to equilibrium critical behavior along a marginal direction. Recent studies propose that systems with continuously varying critical exponents form a ``superuniversality class''  \cite{Vicari, Mukherjee2023} when some of their scaling functions match with those obtained for a parent universality class. Since in the $\w \to \infty$ limit particles become passive resulting in an equilibrium \pst~at $\Jcs = 2\ln(1 + \sqrt{2}),~\rc = 1/2$ [Eq.~\eqref{eq: Jcs}], belonging to the \iuc,  we anticipate that the parent universality class of \pst~must be the \iuc. Indeed, we find that, all along the critical line, the scaling function $B_L$ of the geometric transition as a function of $\xi_2/L,$ matches well with the same obtained for the percolation transition of the lattice gas model belonging to $Z_2P$ universality class. This implies that the  \pst~in the IRTP model falls in the $Z_2$ superuniversality class.     
	
	\section{Conclusion and Outlook \label{sec: VIII}}
	
	In conclusion, we have studied a system of IRTPs on a square lattice that interact via nearest-neighbor attraction and obey steric constraints, forbidding multiple occupancy. This minimal model captures rich phase behavior, exhibiting phase separation transitions driven by motility, interaction, or a combination of both.
	
	To study the critical behavior of the MIPS, we exploit the fact that a critical phase separation transition in 2D is associated with a simultaneous percolation transition occurring there. Although the fact is well established in equilibrium systems like lattice gas models in 2D, its validity in active-matter systems is something we verify explicitly in this article. We primarily study the percolation transition in the IRTP model on a square lattice, determine the critical exponents from the finite-size scaling analysis, and employ the exponent relations given by Eq.~\eqref{eq: MIPS_Perc_Relation} to predict the corresponding critical exponents of MIPS transition. We then verify the predicted exponents in selected cases by examining the MIPS transition, calculating the order parameter $\tilde{\phi}$ [Eq.~\eqref{eq: MIPS_Order_Par}] from the sub-box densities on a rectangular lattice of size $6\ell \times 2\ell$.
	
	We find that the critical exponents of MIPS phase transition vary continuously; however, a scaling function, Binder cumulant as a function of $\xi_2/L$, remains invariant all along the critical line. For the percolation study, the scaling function is found to be identical to that obtained for the interaction-percolation transition in the CLG model, indicating that the percolation of RTPs belongs to the $Z_2P$ superuniversality class and that the corresponding MIPS transition falls within the Ising (or $Z_2$) superuniversality class.
	
	Moreover, our study also illustrates the role of attractive interaction in the MIPS transition. Surprisingly, when combined with an explicit attractive interaction, the overall effective attraction on RTPs appears to weaken. We believe that the attraction induces inhomogeneity in the speed of particles: The effective speed of the RTPs is higher along directions where local density is higher. This inhomogeneity in speed is known to produce effective repulsion \cite{Urna2025}. To understand these phenomena more deeply, we need to analyze systems with a small number of particles, where exact steady states may be accessible.
	
	Importantly, lattice-based models such as the one studied here offer a significant advantage over their continuum counterparts: In certain limiting cases, they usually reduce to equilibrium models. This connection helps anchor the critical behavior of active systems to well-established results from equilibrium statistical mechanics, thereby enabling analytical insights that complement and extend numerical findings.
	
	
	\section*{Acknowledgment}
	We acknowledge the Kepler Computing Facility, maintained by DPS, IISER Kolkata, for providing the computational support. SM gratefully acknowledges financial support through a National Postdoctoral Fellowship from the Anusandhan National Research Foundation (ANRF), Department of Science and Technology (DST), Government of India, under Project  No. PDF/2023/002952. PKM acknowledges the financial support provided by ANRF, Science and Engineering Research Board (SERB), DST, Government of India under Grant No. MTR/2023/000644. 
	
	\newpage
	\bibliography{reference}

\begin{thebibliography}{73}%
\makeatletter
\providecommand \@ifxundefined [1]{%
 \@ifx{#1\undefined}
}%
\providecommand \@ifnum [1]{%
 \ifnum #1\expandafter \@firstoftwo
 \else \expandafter \@secondoftwo
 \fi
}%
\providecommand \@ifx [1]{%
 \ifx #1\expandafter \@firstoftwo
 \else \expandafter \@secondoftwo
 \fi
}%
\providecommand \natexlab [1]{#1}%
\providecommand \enquote  [1]{``#1''}%
\providecommand \bibnamefont  [1]{#1}%
\providecommand \bibfnamefont [1]{#1}%
\providecommand \citenamefont [1]{#1}%
\providecommand \href@noop [0]{\@secondoftwo}%
\providecommand \href [0]{\begingroup \@sanitize@url \@href}%
\providecommand \@href[1]{\@@startlink{#1}\@@href}%
\providecommand \@@href[1]{\endgroup#1\@@endlink}%
\providecommand \@sanitize@url [0]{\catcode `\\12\catcode `\$12\catcode
  `\&12\catcode `\#12\catcode `\^12\catcode `\_12\catcode `\%12\relax}%
\providecommand \@@startlink[1]{}%
\providecommand \@@endlink[0]{}%
\providecommand \url  [0]{\begingroup\@sanitize@url \@url }%
\providecommand \@url [1]{\endgroup\@href {#1}{\urlprefix }}%
\providecommand \urlprefix  [0]{URL }%
\providecommand \Eprint [0]{\href }%
\providecommand \doibase [0]{https://doi.org/}%
\providecommand \selectlanguage [0]{\@gobble}%
\providecommand \bibinfo  [0]{\@secondoftwo}%
\providecommand \bibfield  [0]{\@secondoftwo}%
\providecommand \translation [1]{[#1]}%
\providecommand \BibitemOpen [0]{}%
\providecommand \bibitemStop [0]{}%
\providecommand \bibitemNoStop [0]{.\EOS\space}%
\providecommand \EOS [0]{\spacefactor3000\relax}%
\providecommand \BibitemShut  [1]{\csname bibitem#1\endcsname}%
\let\auto@bib@innerbib\@empty
\bibitem [{\citenamefont {Ramaswamy}(2010)}]{Ramaswamy2010}%
  \BibitemOpen
  \bibfield  {author} {\bibinfo {author} {\bibfnamefont {S.}~\bibnamefont
  {Ramaswamy}},\ }\bibfield  {title} {\bibinfo {title} {The mechanics and
  statistics of active matter},\ }\href
  {http://dx.doi.org/10.1146/annurev-conmatphys-070909-104101} {\bibfield
  {journal} {\bibinfo  {journal} {Annu. Rev. Condens. Matter Phys.}\ }\textbf
  {\bibinfo {volume} {1}},\ \bibinfo {pages} {323} (\bibinfo {year}
  {2010})}\BibitemShut {NoStop}%
\bibitem [{\citenamefont {Cates}(2012)}]{Cates2012}%
  \BibitemOpen
  \bibfield  {author} {\bibinfo {author} {\bibfnamefont {M.~E.}\ \bibnamefont
  {Cates}},\ }\bibfield  {title} {\bibinfo {title} {Diffusive transport without
  detailed balance in motile bacteria: does microbiology need statistical
  physics?},\ }\href {http://dx.doi.org/10.1088/0034-4885/75/4/042601}
  {\bibfield  {journal} {\bibinfo  {journal} {Rep. Prog. Phys.}\ }\textbf
  {\bibinfo {volume} {75}},\ \bibinfo {pages} {042601} (\bibinfo {year}
  {2012})}\BibitemShut {NoStop}%
\bibitem [{\citenamefont {Marchetti}\ \emph {et~al.}(2013)\citenamefont
  {Marchetti}, \citenamefont {Joanny}, \citenamefont {Ramaswamy}, \citenamefont
  {Liverpool}, \citenamefont {Prost}, \citenamefont {Rao},\ and\ \citenamefont
  {Simha}}]{Marchetti2013}%
  \BibitemOpen
  \bibfield  {author} {\bibinfo {author} {\bibfnamefont {M.~C.}\ \bibnamefont
  {Marchetti}}, \bibinfo {author} {\bibfnamefont {J.~F.}\ \bibnamefont
  {Joanny}}, \bibinfo {author} {\bibfnamefont {S.}~\bibnamefont {Ramaswamy}},
  \bibinfo {author} {\bibfnamefont {T.~B.}\ \bibnamefont {Liverpool}}, \bibinfo
  {author} {\bibfnamefont {J.}~\bibnamefont {Prost}}, \bibinfo {author}
  {\bibfnamefont {M.}~\bibnamefont {Rao}},\ and\ \bibinfo {author}
  {\bibfnamefont {R.~A.}\ \bibnamefont {Simha}},\ }\bibfield  {title} {\bibinfo
  {title} {Hydrodynamics of soft active matter},\ }\href
  {https://doi.org/10.1103/revmodphys.85.1143} {\bibfield  {journal} {\bibinfo
  {journal} {Rev. Mod. Phys.}\ }\textbf {\bibinfo {volume} {85}},\ \bibinfo
  {pages} {1143} (\bibinfo {year} {2013})}\BibitemShut {NoStop}%
\bibitem [{\citenamefont {Ballerini}\ \emph {et~al.}(2008)\citenamefont
  {Ballerini}, \citenamefont {Cabibbo}, \citenamefont {Candelier},
  \citenamefont {Cavagna}, \citenamefont {Cisbani}, \citenamefont {Giardina},
  \citenamefont {Lecomte}, \citenamefont {Orlandi}, \citenamefont {Parisi},
  \citenamefont {Procaccini}, \citenamefont {Viale},\ and\ \citenamefont
  {Zdravkovic}}]{Ballerini2008}%
  \BibitemOpen
  \bibfield  {author} {\bibinfo {author} {\bibfnamefont {M.}~\bibnamefont
  {Ballerini}}, \bibinfo {author} {\bibfnamefont {N.}~\bibnamefont {Cabibbo}},
  \bibinfo {author} {\bibfnamefont {R.}~\bibnamefont {Candelier}}, \bibinfo
  {author} {\bibfnamefont {A.}~\bibnamefont {Cavagna}}, \bibinfo {author}
  {\bibfnamefont {E.}~\bibnamefont {Cisbani}}, \bibinfo {author} {\bibfnamefont
  {I.}~\bibnamefont {Giardina}}, \bibinfo {author} {\bibfnamefont
  {V.}~\bibnamefont {Lecomte}}, \bibinfo {author} {\bibfnamefont
  {A.}~\bibnamefont {Orlandi}}, \bibinfo {author} {\bibfnamefont
  {G.}~\bibnamefont {Parisi}}, \bibinfo {author} {\bibfnamefont
  {A.}~\bibnamefont {Procaccini}}, \bibinfo {author} {\bibfnamefont
  {M.}~\bibnamefont {Viale}},\ and\ \bibinfo {author} {\bibfnamefont
  {V.}~\bibnamefont {Zdravkovic}},\ }\bibfield  {title} {\bibinfo {title}
  {Interaction ruling animal collective behavior depends on topological rather
  than metric distance: Evidence from a field study},\ }\href
  {https://doi.org/10.1073/pnas.0711437105} {\bibfield  {journal} {\bibinfo
  {journal} {PNAS}\ }\textbf {\bibinfo {volume} {105}},\ \bibinfo {pages}
  {1232} (\bibinfo {year} {2008})}\BibitemShut {NoStop}%
\bibitem [{\citenamefont {Katz}\ \emph {et~al.}(2011)\citenamefont {Katz},
  \citenamefont {Tunstrøm}, \citenamefont {Ioannou}, \citenamefont {Huepe},\
  and\ \citenamefont {Couzin}}]{Katz2011}%
  \BibitemOpen
  \bibfield  {author} {\bibinfo {author} {\bibfnamefont {Y.}~\bibnamefont
  {Katz}}, \bibinfo {author} {\bibfnamefont {K.}~\bibnamefont {Tunstrøm}},
  \bibinfo {author} {\bibfnamefont {C.~C.}\ \bibnamefont {Ioannou}}, \bibinfo
  {author} {\bibfnamefont {C.}~\bibnamefont {Huepe}},\ and\ \bibinfo {author}
  {\bibfnamefont {I.~D.}\ \bibnamefont {Couzin}},\ }\bibfield  {title}
  {\bibinfo {title} {Inferring the structure and dynamics of interactions in
  schooling fish},\ }\href {https://doi.org/10.1073/pnas.1107583108} {\bibfield
   {journal} {\bibinfo  {journal} {PNAS}\ }\textbf {\bibinfo {volume} {108}},\
  \bibinfo {pages} {18720} (\bibinfo {year} {2011})}\BibitemShut {NoStop}%
\bibitem [{\citenamefont {Schaller}\ \emph {et~al.}(2010)\citenamefont
  {Schaller}, \citenamefont {Weber}, \citenamefont {Semmrich}, \citenamefont
  {Frey},\ and\ \citenamefont {Bausch}}]{Schaller2010}%
  \BibitemOpen
  \bibfield  {author} {\bibinfo {author} {\bibfnamefont {V.}~\bibnamefont
  {Schaller}}, \bibinfo {author} {\bibfnamefont {C.}~\bibnamefont {Weber}},
  \bibinfo {author} {\bibfnamefont {C.}~\bibnamefont {Semmrich}}, \bibinfo
  {author} {\bibfnamefont {E.}~\bibnamefont {Frey}},\ and\ \bibinfo {author}
  {\bibfnamefont {A.~R.}\ \bibnamefont {Bausch}},\ }\bibfield  {title}
  {\bibinfo {title} {Polar patterns of driven filaments},\ }\href
  {https://doi.org/10.1038/nature09312} {\bibfield  {journal} {\bibinfo
  {journal} {Nat. Commun.}\ }\textbf {\bibinfo {volume} {467}},\ \bibinfo
  {pages} {73} (\bibinfo {year} {2010})}\BibitemShut {NoStop}%
\bibitem [{\citenamefont {Sumino}\ \emph {et~al.}(2012)\citenamefont {Sumino},
  \citenamefont {Nagai}, \citenamefont {Shitaka}, \citenamefont {Tanaka},
  \citenamefont {Yoshikawa}, \citenamefont {Chaté},\ and\ \citenamefont
  {Oiwa}}]{Sumino2012}%
  \BibitemOpen
  \bibfield  {author} {\bibinfo {author} {\bibfnamefont {Y.}~\bibnamefont
  {Sumino}}, \bibinfo {author} {\bibfnamefont {K.~H.}\ \bibnamefont {Nagai}},
  \bibinfo {author} {\bibfnamefont {Y.}~\bibnamefont {Shitaka}}, \bibinfo
  {author} {\bibfnamefont {D.}~\bibnamefont {Tanaka}}, \bibinfo {author}
  {\bibfnamefont {K.}~\bibnamefont {Yoshikawa}}, \bibinfo {author}
  {\bibfnamefont {H.}~\bibnamefont {Chaté}},\ and\ \bibinfo {author}
  {\bibfnamefont {K.}~\bibnamefont {Oiwa}},\ }\bibfield  {title} {\bibinfo
  {title} {Large-scale vortex lattice emerging from collectively moving
  microtubules},\ }\href {https://doi.org/10.1038/nature10874} {\bibfield
  {journal} {\bibinfo  {journal} {Nat. Commun.}\ }\textbf {\bibinfo {volume}
  {483}},\ \bibinfo {pages} {448} (\bibinfo {year} {2012})}\BibitemShut
  {NoStop}%
\bibitem [{\citenamefont {Theurkauff}\ \emph {et~al.}(2012)\citenamefont
  {Theurkauff}, \citenamefont {Cottin-Bizonne}, \citenamefont {Palacci},
  \citenamefont {Ybert},\ and\ \citenamefont {Bocquet}}]{Theurkauff2012}%
  \BibitemOpen
  \bibfield  {author} {\bibinfo {author} {\bibfnamefont {I.}~\bibnamefont
  {Theurkauff}}, \bibinfo {author} {\bibfnamefont {C.}~\bibnamefont
  {Cottin-Bizonne}}, \bibinfo {author} {\bibfnamefont {J.}~\bibnamefont
  {Palacci}}, \bibinfo {author} {\bibfnamefont {C.}~\bibnamefont {Ybert}},\
  and\ \bibinfo {author} {\bibfnamefont {L.}~\bibnamefont {Bocquet}},\
  }\bibfield  {title} {\bibinfo {title} {Dynamic clustering in active colloidal
  suspensions with chemical signaling},\ }\href
  {http://dx.doi.org/10.1103/PhysRevLett.108.268303} {\bibfield  {journal}
  {\bibinfo  {journal} {Phys. Rev. Lett.}\ }\textbf {\bibinfo {volume} {108}},\
  \bibinfo {pages} {268303} (\bibinfo {year} {2012})}\BibitemShut {NoStop}%
\bibitem [{\citenamefont {Buttinoni}\ \emph {et~al.}(2013)\citenamefont
  {Buttinoni}, \citenamefont {Bialké}, \citenamefont {K\"{u}mmel},
  \citenamefont {L\"{o}wen}, \citenamefont {Bechinger},\ and\ \citenamefont
  {Speck}}]{Buttinoni2013}%
  \BibitemOpen
  \bibfield  {author} {\bibinfo {author} {\bibfnamefont {I.}~\bibnamefont
  {Buttinoni}}, \bibinfo {author} {\bibfnamefont {J.}~\bibnamefont {Bialké}},
  \bibinfo {author} {\bibfnamefont {F.}~\bibnamefont {K\"{u}mmel}}, \bibinfo
  {author} {\bibfnamefont {H.}~\bibnamefont {L\"{o}wen}}, \bibinfo {author}
  {\bibfnamefont {C.}~\bibnamefont {Bechinger}},\ and\ \bibinfo {author}
  {\bibfnamefont {T.}~\bibnamefont {Speck}},\ }\bibfield  {title} {\bibinfo
  {title} {Dynamical clustering and phase separation in suspensions of
  self-propelled colloidal particles},\ }\href
  {http://dx.doi.org/10.1103/PhysRevLett.110.238301} {\bibfield  {journal}
  {\bibinfo  {journal} {Phys. Rev. Lett.}\ }\textbf {\bibinfo {volume} {110}},\
  \bibinfo {pages} {238301} (\bibinfo {year} {2013})}\BibitemShut {NoStop}%
\bibitem [{\citenamefont {Palacci}\ \emph {et~al.}(2013)\citenamefont
  {Palacci}, \citenamefont {Sacanna}, \citenamefont {Steinberg}, \citenamefont
  {Pine},\ and\ \citenamefont {Chaikin}}]{Palacci2013}%
  \BibitemOpen
  \bibfield  {author} {\bibinfo {author} {\bibfnamefont {J.}~\bibnamefont
  {Palacci}}, \bibinfo {author} {\bibfnamefont {S.}~\bibnamefont {Sacanna}},
  \bibinfo {author} {\bibfnamefont {A.~P.}\ \bibnamefont {Steinberg}}, \bibinfo
  {author} {\bibfnamefont {D.~J.}\ \bibnamefont {Pine}},\ and\ \bibinfo
  {author} {\bibfnamefont {P.~M.}\ \bibnamefont {Chaikin}},\ }\bibfield
  {title} {\bibinfo {title} {Living crystals of light-activated colloidal
  surfers},\ }\href {https://doi.org/10.1126/science.1230020} {\bibfield
  {journal} {\bibinfo  {journal} {Science}\ }\textbf {\bibinfo {volume}
  {339}},\ \bibinfo {pages} {936} (\bibinfo {year} {2013})}\BibitemShut
  {NoStop}%
\bibitem [{\citenamefont {Bricard}\ \emph {et~al.}(2013)\citenamefont
  {Bricard}, \citenamefont {Caussin}, \citenamefont {Desreumaux}, \citenamefont
  {Dauchot},\ and\ \citenamefont {Bartolo}}]{Bricard2013}%
  \BibitemOpen
  \bibfield  {author} {\bibinfo {author} {\bibfnamefont {A.}~\bibnamefont
  {Bricard}}, \bibinfo {author} {\bibfnamefont {J.-B.}\ \bibnamefont
  {Caussin}}, \bibinfo {author} {\bibfnamefont {N.}~\bibnamefont {Desreumaux}},
  \bibinfo {author} {\bibfnamefont {O.}~\bibnamefont {Dauchot}},\ and\ \bibinfo
  {author} {\bibfnamefont {D.}~\bibnamefont {Bartolo}},\ }\bibfield  {title}
  {\bibinfo {title} {Emergence of macroscopic directed motion in populations of
  motile colloids},\ }\href {https://doi.org/10.1038/nature12673} {\bibfield
  {journal} {\bibinfo  {journal} {Nat. Commun.}\ }\textbf {\bibinfo {volume}
  {503}},\ \bibinfo {pages} {95} (\bibinfo {year} {2013})}\BibitemShut
  {NoStop}%
\bibitem [{\citenamefont {Schnitzer}(1993)}]{Schnitzer1993}%
  \BibitemOpen
  \bibfield  {author} {\bibinfo {author} {\bibfnamefont {M.~J.}\ \bibnamefont
  {Schnitzer}},\ }\bibfield  {title} {\bibinfo {title} {Theory of continuum
  random walks and application to chemotaxis},\ }\href
  {https://doi.org/10.1103/physreve.48.2553} {\bibfield  {journal} {\bibinfo
  {journal} {Phys. Rev. E}\ }\textbf {\bibinfo {volume} {48}},\ \bibinfo
  {pages} {2553} (\bibinfo {year} {1993})}\BibitemShut {NoStop}%
\bibitem [{\citenamefont {Berg}(2004)}]{H.C.Berg2004}%
  \BibitemOpen
  \bibfield  {author} {\bibinfo {author} {\bibfnamefont {H.~C.}\ \bibnamefont
  {Berg}},\ }\href {https://doi.org/10.1007/b97370} {\emph {\bibinfo {title}
  {E. coli in Motion}}}\ (\bibinfo  {publisher} {Springer New York},\ \bibinfo
  {year} {2004})\BibitemShut {NoStop}%
\bibitem [{\citenamefont {Romanczuk}\ \emph {et~al.}(2012)\citenamefont
  {Romanczuk}, \citenamefont {B\"{a}r}, \citenamefont {Ebeling}, \citenamefont
  {Lindner},\ and\ \citenamefont {Schimansky-Geier}}]{Romanczuk2012}%
  \BibitemOpen
  \bibfield  {author} {\bibinfo {author} {\bibfnamefont {P.}~\bibnamefont
  {Romanczuk}}, \bibinfo {author} {\bibfnamefont {M.}~\bibnamefont {B\"{a}r}},
  \bibinfo {author} {\bibfnamefont {W.}~\bibnamefont {Ebeling}}, \bibinfo
  {author} {\bibfnamefont {B.}~\bibnamefont {Lindner}},\ and\ \bibinfo {author}
  {\bibfnamefont {L.}~\bibnamefont {Schimansky-Geier}},\ }\bibfield  {title}
  {\bibinfo {title} {Active {Brownian} particles: From individual to collective
  stochastic dynamics},\ }\href {https://doi.org/10.1140/epjst/e2012-01529-y}
  {\bibfield  {journal} {\bibinfo  {journal} {Eur. Phys. J. Spec. Top.}\
  }\textbf {\bibinfo {volume} {202}},\ \bibinfo {pages} {1} (\bibinfo {year}
  {2012})}\BibitemShut {NoStop}%
\bibitem [{\citenamefont {Polin}\ \emph {et~al.}(2009)\citenamefont {Polin},
  \citenamefont {Tuval}, \citenamefont {Drescher}, \citenamefont {Gollub},\
  and\ \citenamefont {Goldstein}}]{Polin2009}%
  \BibitemOpen
  \bibfield  {author} {\bibinfo {author} {\bibfnamefont {M.}~\bibnamefont
  {Polin}}, \bibinfo {author} {\bibfnamefont {I.}~\bibnamefont {Tuval}},
  \bibinfo {author} {\bibfnamefont {K.}~\bibnamefont {Drescher}}, \bibinfo
  {author} {\bibfnamefont {J.~P.}\ \bibnamefont {Gollub}},\ and\ \bibinfo
  {author} {\bibfnamefont {R.~E.}\ \bibnamefont {Goldstein}},\ }\bibfield
  {title} {\bibinfo {title} {Chlamydomonas swims with two “gears” in a
  eukaryotic version of run-and-tumble locomotion},\ }\href
  {https://doi.org/10.1126/science.1172667} {\bibfield  {journal} {\bibinfo
  {journal} {Science}\ }\textbf {\bibinfo {volume} {325}},\ \bibinfo {pages}
  {487} (\bibinfo {year} {2009})}\BibitemShut {NoStop}%
\bibitem [{\citenamefont {Cates}\ and\ \citenamefont
  {Tailleur}(2013)}]{Cates2013}%
  \BibitemOpen
  \bibfield  {author} {\bibinfo {author} {\bibfnamefont {M.~E.}\ \bibnamefont
  {Cates}}\ and\ \bibinfo {author} {\bibfnamefont {J.}~\bibnamefont
  {Tailleur}},\ }\bibfield  {title} {\bibinfo {title} {When are active
  {Brownian} particles and run-and-tumble particles equivalent? consequences
  for motility-induced phase separation},\ }\href
  {https://doi.org/10.1209/0295-5075/101/20010} {\bibfield  {journal} {\bibinfo
   {journal} {Europhys. Lett.}\ }\textbf {\bibinfo {volume} {101}},\ \bibinfo
  {pages} {20010} (\bibinfo {year} {2013})}\BibitemShut {NoStop}%
\bibitem [{\citenamefont {Tailleur}\ and\ \citenamefont
  {Cates}(2009)}]{Tailleur2009}%
  \BibitemOpen
  \bibfield  {author} {\bibinfo {author} {\bibfnamefont {J.}~\bibnamefont
  {Tailleur}}\ and\ \bibinfo {author} {\bibfnamefont {M.~E.}\ \bibnamefont
  {Cates}},\ }\bibfield  {title} {\bibinfo {title} {Sedimentation, trapping,
  and rectification of dilute bacteria},\ }\href
  {https://doi.org/10.1209/0295-5075/86/60002} {\bibfield  {journal} {\bibinfo
  {journal} {Europhys. Lett.}\ }\textbf {\bibinfo {volume} {86}},\ \bibinfo
  {pages} {60002} (\bibinfo {year} {2009})}\BibitemShut {NoStop}%
\bibitem [{\citenamefont {Enculescu}\ and\ \citenamefont
  {Stark}(2011)}]{Enculescu2011}%
  \BibitemOpen
  \bibfield  {author} {\bibinfo {author} {\bibfnamefont {M.}~\bibnamefont
  {Enculescu}}\ and\ \bibinfo {author} {\bibfnamefont {H.}~\bibnamefont
  {Stark}},\ }\bibfield  {title} {\bibinfo {title} {Active colloidal
  suspensions exhibit polar order under gravity},\ }\href
  {http://dx.doi.org/10.1103/PhysRevLett.107.058301} {\bibfield  {journal}
  {\bibinfo  {journal} {Phys. Rev. Lett.}\ }\textbf {\bibinfo {volume} {107}},\
  \bibinfo {pages} {058301} (\bibinfo {year} {2011})}\BibitemShut {NoStop}%
\bibitem [{\citenamefont {Lee}(2013)}]{Lee2013}%
  \BibitemOpen
  \bibfield  {author} {\bibinfo {author} {\bibfnamefont {C.~F.}\ \bibnamefont
  {Lee}},\ }\bibfield  {title} {\bibinfo {title} {Active particles under
  confinement: aggregation at the wall and gradient formation inside a
  channel},\ }\href {https://doi.org/10.1088/1367-2630/15/5/055007} {\bibfield
  {journal} {\bibinfo  {journal} {New J. Phys.}\ }\textbf {\bibinfo {volume}
  {15}},\ \bibinfo {pages} {055007} (\bibinfo {year} {2013})}\BibitemShut
  {NoStop}%
\bibitem [{\citenamefont {Slowman}\ \emph {et~al.}(2016)\citenamefont
  {Slowman}, \citenamefont {Evans},\ and\ \citenamefont
  {Blythe}}]{Slowman2016}%
  \BibitemOpen
  \bibfield  {author} {\bibinfo {author} {\bibfnamefont {A.}~\bibnamefont
  {Slowman}}, \bibinfo {author} {\bibfnamefont {M.}~\bibnamefont {Evans}},\
  and\ \bibinfo {author} {\bibfnamefont {R.}~\bibnamefont {Blythe}},\
  }\bibfield  {title} {\bibinfo {title} {Jamming and attraction of interacting
  run-and-tumble random walkers},\ }\href
  {http://dx.doi.org/10.1103/PhysRevLett.116.218101} {\bibfield  {journal}
  {\bibinfo  {journal} {Phys. Rev. Lett.}\ }\textbf {\bibinfo {volume} {116}},\
  \bibinfo {pages} {218101} (\bibinfo {year} {2016})}\BibitemShut {NoStop}%
\bibitem [{\citenamefont {Slowman}\ \emph {et~al.}(2017)\citenamefont
  {Slowman}, \citenamefont {Evans},\ and\ \citenamefont
  {Blythe}}]{Slowman2017}%
  \BibitemOpen
  \bibfield  {author} {\bibinfo {author} {\bibfnamefont {A.~B.}\ \bibnamefont
  {Slowman}}, \bibinfo {author} {\bibfnamefont {M.~R.}\ \bibnamefont {Evans}},\
  and\ \bibinfo {author} {\bibfnamefont {R.~A.}\ \bibnamefont {Blythe}},\
  }\bibfield  {title} {\bibinfo {title} {Exact solution of two interacting
  run-and-tumble random walkers with finite tumble duration},\ }\href
  {https://doi.org/10.1088/1751-8121/aa80af} {\bibfield  {journal} {\bibinfo
  {journal} {J. Phys. A: Math. Theor.}\ }\textbf {\bibinfo {volume} {50}},\
  \bibinfo {pages} {375601} (\bibinfo {year} {2017})}\BibitemShut {NoStop}%
\bibitem [{\citenamefont {Reichhardt}\ and\ \citenamefont
  {Reichhardt}(2017)}]{Reichhardt2017}%
  \BibitemOpen
  \bibfield  {author} {\bibinfo {author} {\bibfnamefont {C.~O.}\ \bibnamefont
  {Reichhardt}}\ and\ \bibinfo {author} {\bibfnamefont {C.}~\bibnamefont
  {Reichhardt}},\ }\bibfield  {title} {\bibinfo {title} {Ratchet effects in
  active matter systems},\ }\href
  {https://doi.org/10.1146/annurev-conmatphys-031016-025522} {\bibfield
  {journal} {\bibinfo  {journal} {Annu. Rev. Condens. Matter Phys.}\ }\textbf
  {\bibinfo {volume} {8}},\ \bibinfo {pages} {51} (\bibinfo {year}
  {2017})}\BibitemShut {NoStop}%
\bibitem [{\citenamefont {Tailleur}\ and\ \citenamefont
  {Cates}(2008)}]{Tailleur2008}%
  \BibitemOpen
  \bibfield  {author} {\bibinfo {author} {\bibfnamefont {J.}~\bibnamefont
  {Tailleur}}\ and\ \bibinfo {author} {\bibfnamefont {M.~E.}\ \bibnamefont
  {Cates}},\ }\bibfield  {title} {\bibinfo {title} {Statistical mechanics of
  interacting run-and-tumble bacteria},\ }\href
  {http://dx.doi.org/10.1103/PhysRevLett.100.218103} {\bibfield  {journal}
  {\bibinfo  {journal} {Phys. Rev. Lett.}\ }\textbf {\bibinfo {volume} {100}},\
  \bibinfo {pages} {218103} (\bibinfo {year} {2008})}\BibitemShut {NoStop}%
\bibitem [{\citenamefont {Redner}\ \emph
  {et~al.}(2013{\natexlab{a}})\citenamefont {Redner}, \citenamefont {Hagan},\
  and\ \citenamefont {Baskaran}}]{Redner2013}%
  \BibitemOpen
  \bibfield  {author} {\bibinfo {author} {\bibfnamefont {G.~S.}\ \bibnamefont
  {Redner}}, \bibinfo {author} {\bibfnamefont {M.~F.}\ \bibnamefont {Hagan}},\
  and\ \bibinfo {author} {\bibfnamefont {A.}~\bibnamefont {Baskaran}},\
  }\bibfield  {title} {\bibinfo {title} {Structure and dynamics of a
  phase-separating active colloidal fluid},\ }\href
  {http://dx.doi.org/10.1103/PhysRevLett.110.055701} {\bibfield  {journal}
  {\bibinfo  {journal} {Phys. Rev. Lett.}\ }\textbf {\bibinfo {volume} {110}},\
  \bibinfo {pages} {055701} (\bibinfo {year} {2013}{\natexlab{a}})}\BibitemShut
  {NoStop}%
\bibitem [{\citenamefont {Redner}\ \emph
  {et~al.}(2013{\natexlab{b}})\citenamefont {Redner}, \citenamefont
  {Baskaran},\ and\ \citenamefont {Hagan}}]{Redner2013a}%
  \BibitemOpen
  \bibfield  {author} {\bibinfo {author} {\bibfnamefont {G.~S.}\ \bibnamefont
  {Redner}}, \bibinfo {author} {\bibfnamefont {A.}~\bibnamefont {Baskaran}},\
  and\ \bibinfo {author} {\bibfnamefont {M.~F.}\ \bibnamefont {Hagan}},\
  }\bibfield  {title} {\bibinfo {title} {Reentrant phase behavior in active
  colloids with attraction},\ }\href
  {https://doi.org/10.1103/PhysRevE.88.012305} {\bibfield  {journal} {\bibinfo
  {journal} {Phys. Rev. E}\ }\textbf {\bibinfo {volume} {88}},\ \bibinfo
  {pages} {012305} (\bibinfo {year} {2013}{\natexlab{b}})}\BibitemShut
  {NoStop}%
\bibitem [{\citenamefont {Stenhammar}\ \emph {et~al.}(2013)\citenamefont
  {Stenhammar}, \citenamefont {Tiribocchi}, \citenamefont {Allen},
  \citenamefont {Marenduzzo},\ and\ \citenamefont {Cates}}]{Stenhammar2013}%
  \BibitemOpen
  \bibfield  {author} {\bibinfo {author} {\bibfnamefont {J.}~\bibnamefont
  {Stenhammar}}, \bibinfo {author} {\bibfnamefont {A.}~\bibnamefont
  {Tiribocchi}}, \bibinfo {author} {\bibfnamefont {R.~J.}\ \bibnamefont
  {Allen}}, \bibinfo {author} {\bibfnamefont {D.}~\bibnamefont {Marenduzzo}},\
  and\ \bibinfo {author} {\bibfnamefont {M.~E.}\ \bibnamefont {Cates}},\
  }\bibfield  {title} {\bibinfo {title} {Continuum theory of phase separation
  kinetics for active {Brownian} particles},\ }\href
  {http://dx.doi.org/10.1103/PhysRevLett.111.145702} {\bibfield  {journal}
  {\bibinfo  {journal} {Phys. Rev. Lett.}\ }\textbf {\bibinfo {volume} {111}},\
  \bibinfo {pages} {145702} (\bibinfo {year} {2013})}\BibitemShut {NoStop}%
\bibitem [{\citenamefont {Cates}\ and\ \citenamefont
  {Tailleur}(2015)}]{Cates2015}%
  \BibitemOpen
  \bibfield  {author} {\bibinfo {author} {\bibfnamefont {M.~E.}\ \bibnamefont
  {Cates}}\ and\ \bibinfo {author} {\bibfnamefont {J.}~\bibnamefont
  {Tailleur}},\ }\bibfield  {title} {\bibinfo {title} {Motility-induced phase
  separation},\ }\href
  {https://doi.org/10.1146/annurev-conmatphys-031214-014710} {\bibfield
  {journal} {\bibinfo  {journal} {Annu. Rev. Condens. Matter Phys.}\ }\textbf
  {\bibinfo {volume} {6}},\ \bibinfo {pages} {219} (\bibinfo {year}
  {2015})}\BibitemShut {NoStop}%
\bibitem [{\citenamefont {Patch}\ \emph {et~al.}(2017)\citenamefont {Patch},
  \citenamefont {Yllanes},\ and\ \citenamefont {Marchetti}}]{Patch2017}%
  \BibitemOpen
  \bibfield  {author} {\bibinfo {author} {\bibfnamefont {A.}~\bibnamefont
  {Patch}}, \bibinfo {author} {\bibfnamefont {D.}~\bibnamefont {Yllanes}},\
  and\ \bibinfo {author} {\bibfnamefont {M.~C.}\ \bibnamefont {Marchetti}},\
  }\bibfield  {title} {\bibinfo {title} {Kinetics of motility-induced phase
  separation and swim pressure},\ }\href
  {http://dx.doi.org/10.1103/PhysRevE.95.012601} {\bibfield  {journal}
  {\bibinfo  {journal} {Phys. Rev. E}\ }\textbf {\bibinfo {volume} {95}},\
  \bibinfo {pages} {012601} (\bibinfo {year} {2017})}\BibitemShut {NoStop}%
\bibitem [{\citenamefont {Fily}\ and\ \citenamefont
  {Marchetti}(2012)}]{Fily2012}%
  \BibitemOpen
  \bibfield  {author} {\bibinfo {author} {\bibfnamefont {Y.}~\bibnamefont
  {Fily}}\ and\ \bibinfo {author} {\bibfnamefont {M.~C.}\ \bibnamefont
  {Marchetti}},\ }\bibfield  {title} {\bibinfo {title} {Athermal phase
  separation of self-propelled particles with no alignment},\ }\href
  {http://dx.doi.org/10.1103/PhysRevLett.108.235702} {\bibfield  {journal}
  {\bibinfo  {journal} {Phys. Rev. Lett.}\ }\textbf {\bibinfo {volume} {108}},\
  \bibinfo {pages} {235702} (\bibinfo {year} {2012})}\BibitemShut {NoStop}%
\bibitem [{\citenamefont {Bialké}\ \emph {et~al.}(2013)\citenamefont
  {Bialké}, \citenamefont {L\"{o}wen},\ and\ \citenamefont
  {Speck}}]{Bialk2013}%
  \BibitemOpen
  \bibfield  {author} {\bibinfo {author} {\bibfnamefont {J.}~\bibnamefont
  {Bialké}}, \bibinfo {author} {\bibfnamefont {H.}~\bibnamefont {L\"{o}wen}},\
  and\ \bibinfo {author} {\bibfnamefont {T.}~\bibnamefont {Speck}},\ }\bibfield
   {title} {\bibinfo {title} {Microscopic theory for the phase separation of
  self-propelled repulsive disks},\ }\href
  {https://doi.org/10.1209/0295-5075/103/30008} {\bibfield  {journal} {\bibinfo
   {journal} {Europhys. Lett.}\ }\textbf {\bibinfo {volume} {103}},\ \bibinfo
  {pages} {30008} (\bibinfo {year} {2013})}\BibitemShut {NoStop}%
\bibitem [{\citenamefont {Schweitzer}(2018)}]{Schweitzer2018}%
  \BibitemOpen
  \bibfield  {author} {\bibinfo {author} {\bibfnamefont {F.}~\bibnamefont
  {Schweitzer}},\ }\bibfield  {title} {\bibinfo {title} {An agent-based
  framework of active matter with applications in biological and social
  systems},\ }\href {https://doi.org/10.1088/1361-6404/aaeb63} {\bibfield
  {journal} {\bibinfo  {journal} {Eur. J. Phys.}\ }\textbf {\bibinfo {volume}
  {40}},\ \bibinfo {pages} {014003} (\bibinfo {year} {2018})}\BibitemShut
  {NoStop}%
\bibitem [{\citenamefont {Ziepke}\ \emph {et~al.}(2022)\citenamefont {Ziepke},
  \citenamefont {Maryshev}, \citenamefont {Aranson},\ and\ \citenamefont
  {Frey}}]{Ziepke2022}%
  \BibitemOpen
  \bibfield  {author} {\bibinfo {author} {\bibfnamefont {A.}~\bibnamefont
  {Ziepke}}, \bibinfo {author} {\bibfnamefont {I.}~\bibnamefont {Maryshev}},
  \bibinfo {author} {\bibfnamefont {I.~S.}\ \bibnamefont {Aranson}},\ and\
  \bibinfo {author} {\bibfnamefont {E.}~\bibnamefont {Frey}},\ }\bibfield
  {title} {\bibinfo {title} {Multi-scale organization in communicating active
  matter},\ }\href {http://dx.doi.org/10.1038/s41467-022-34484-2} {\bibfield
  {journal} {\bibinfo  {journal} {Nat. Commun.}\ }\textbf {\bibinfo {volume}
  {13}},\ \bibinfo {pages} {6727} (\bibinfo {year} {2022})}\BibitemShut
  {NoStop}%
\bibitem [{\citenamefont {Thompson}\ \emph {et~al.}(2011)\citenamefont
  {Thompson}, \citenamefont {Tailleur}, \citenamefont {Cates},\ and\
  \citenamefont {Blythe}}]{Thompson2011}%
  \BibitemOpen
  \bibfield  {author} {\bibinfo {author} {\bibfnamefont {A.~G.}\ \bibnamefont
  {Thompson}}, \bibinfo {author} {\bibfnamefont {J.}~\bibnamefont {Tailleur}},
  \bibinfo {author} {\bibfnamefont {M.~E.}\ \bibnamefont {Cates}},\ and\
  \bibinfo {author} {\bibfnamefont {R.~A.}\ \bibnamefont {Blythe}},\ }\bibfield
   {title} {\bibinfo {title} {Lattice models of nonequilibrium bacterial
  dynamics},\ }\href {https://doi.org/10.1088/1742-5468/2011/02/p02029}
  {\bibfield  {journal} {\bibinfo  {journal} {J. Stat. Mech.: Theory Exp.}\
  }\textbf {\bibinfo {volume} {2011}}\bibinfo  {number} { (02)},\ \bibinfo
  {pages} {P02029}}\BibitemShut {NoStop}%
\bibitem [{\citenamefont {Mallmin}\ \emph {et~al.}(2019)\citenamefont
  {Mallmin}, \citenamefont {Blythe},\ and\ \citenamefont
  {Evans}}]{Mallmin2019}%
  \BibitemOpen
\bibfield  {number} {  }\bibfield  {author} {\bibinfo {author} {\bibfnamefont
  {E.}~\bibnamefont {Mallmin}}, \bibinfo {author} {\bibfnamefont {R.~A.}\
  \bibnamefont {Blythe}},\ and\ \bibinfo {author} {\bibfnamefont {M.~R.}\
  \bibnamefont {Evans}},\ }\bibfield  {title} {\bibinfo {title} {Exact spectral
  solution of two interacting run-and-tumble particles on a ring lattice},\
  }\href {https://doi.org/10.1088/1742-5468/aaf631} {\bibfield  {journal}
  {\bibinfo  {journal} {J. Stat. Mech.: Theory Exp.}\ }\textbf {\bibinfo
  {volume} {2019}}\bibinfo  {number} { (1)},\ \bibinfo {pages}
  {013204}}\BibitemShut {NoStop}%
\bibitem [{\citenamefont {Dandekar}\ \emph {et~al.}(2020)\citenamefont
  {Dandekar}, \citenamefont {Chakraborti},\ and\ \citenamefont
  {Rajesh}}]{Dandekar2020}%
  \BibitemOpen
\bibfield  {number} {  }\bibfield  {author} {\bibinfo {author} {\bibfnamefont
  {R.}~\bibnamefont {Dandekar}}, \bibinfo {author} {\bibfnamefont
  {S.}~\bibnamefont {Chakraborti}},\ and\ \bibinfo {author} {\bibfnamefont
  {R.}~\bibnamefont {Rajesh}},\ }\bibfield  {title} {\bibinfo {title} {Hard
  core run and tumble particles on a one-dimensional lattice},\ }\href
  {http://dx.doi.org/10.1103/PhysRevE.102.062111} {\bibfield  {journal}
  {\bibinfo  {journal} {Phys. Rev. E}\ }\textbf {\bibinfo {volume} {102}},\
  \bibinfo {pages} {062111} (\bibinfo {year} {2020})}\BibitemShut {NoStop}%
\bibitem [{\citenamefont {Ray}\ \emph {et~al.}(2024)\citenamefont {Ray},
  \citenamefont {Mukherjee},\ and\ \citenamefont {Mohanty}}]{Ray2024}%
  \BibitemOpen
  \bibfield  {author} {\bibinfo {author} {\bibfnamefont {C.~G.}\ \bibnamefont
  {Ray}}, \bibinfo {author} {\bibfnamefont {I.}~\bibnamefont {Mukherjee}},\
  and\ \bibinfo {author} {\bibfnamefont {P.~K.}\ \bibnamefont {Mohanty}},\
  }\bibfield  {title} {\bibinfo {title} {How motility affects {Ising}
  transitions},\ }\href {https://doi.org/10.1088/1742-5468/ad685b} {\bibfield
  {journal} {\bibinfo  {journal} {J. Stat. Mech.: Theory Exp.}\ }\textbf
  {\bibinfo {volume} {2024}}\bibinfo  {number} { (9)},\ \bibinfo {pages}
  {093207}}\BibitemShut {NoStop}%
\bibitem [{\citenamefont {Kourbane-Houssene}\ \emph {et~al.}(2018)\citenamefont
  {Kourbane-Houssene}, \citenamefont {Erignoux}, \citenamefont {Bodineau},\
  and\ \citenamefont {Tailleur}}]{KourbaneHoussene2018}%
  \BibitemOpen
\bibfield  {number} {  }\bibfield  {author} {\bibinfo {author} {\bibfnamefont
  {M.}~\bibnamefont {Kourbane-Houssene}}, \bibinfo {author} {\bibfnamefont
  {C.}~\bibnamefont {Erignoux}}, \bibinfo {author} {\bibfnamefont
  {T.}~\bibnamefont {Bodineau}},\ and\ \bibinfo {author} {\bibfnamefont
  {J.}~\bibnamefont {Tailleur}},\ }\bibfield  {title} {\bibinfo {title} {Exact
  hydrodynamic description of active lattice gases},\ }\href
  {http://dx.doi.org/10.1103/PhysRevLett.120.268003} {\bibfield  {journal}
  {\bibinfo  {journal} {Phys. Rev. Lett.}\ }\textbf {\bibinfo {volume} {120}},\
  \bibinfo {pages} {268003} (\bibinfo {year} {2018})}\BibitemShut {NoStop}%
\bibitem [{\citenamefont {Sepúlveda}\ and\ \citenamefont
  {Soto}(2016)}]{Seplveda2016}%
  \BibitemOpen
  \bibfield  {author} {\bibinfo {author} {\bibfnamefont {N.}~\bibnamefont
  {Sepúlveda}}\ and\ \bibinfo {author} {\bibfnamefont {R.}~\bibnamefont
  {Soto}},\ }\bibfield  {title} {\bibinfo {title} {Coarsening and clustering in
  run-and-tumble dynamics with short-range exclusion},\ }\href
  {http://dx.doi.org/10.1103/PhysRevE.94.022603} {\bibfield  {journal}
  {\bibinfo  {journal} {Phys. Rev. E}\ }\textbf {\bibinfo {volume} {94}},\
  \bibinfo {pages} {022603} (\bibinfo {year} {2016})}\BibitemShut {NoStop}%
\bibitem [{\citenamefont {Mukherjee}\ \emph {et~al.}(2023)\citenamefont
  {Mukherjee}, \citenamefont {Raghu},\ and\ \citenamefont
  {Mohanty}}]{Mukherjee2023a}%
  \BibitemOpen
  \bibfield  {author} {\bibinfo {author} {\bibfnamefont {I.}~\bibnamefont
  {Mukherjee}}, \bibinfo {author} {\bibfnamefont {A.}~\bibnamefont {Raghu}},\
  and\ \bibinfo {author} {\bibfnamefont {P.~K.}\ \bibnamefont {Mohanty}},\
  }\bibfield  {title} {\bibinfo {title} {Nonexistence of motility induced phase
  separation transition in one dimension},\ }\href
  {http://dx.doi.org/10.21468/SciPostPhys.14.6.165} {\bibfield  {journal}
  {\bibinfo  {journal} {SciPost Phys.}\ }\textbf {\bibinfo {volume} {14}},\
  \bibinfo {pages} {165} (\bibinfo {year} {2023})}\BibitemShut {NoStop}%
\bibitem [{\citenamefont {Dittrich}\ \emph {et~al.}(2021)\citenamefont
  {Dittrich}, \citenamefont {Speck},\ and\ \citenamefont
  {Virnau}}]{Dittrich2021}%
  \BibitemOpen
  \bibfield  {author} {\bibinfo {author} {\bibfnamefont {F.}~\bibnamefont
  {Dittrich}}, \bibinfo {author} {\bibfnamefont {T.}~\bibnamefont {Speck}},\
  and\ \bibinfo {author} {\bibfnamefont {P.}~\bibnamefont {Virnau}},\
  }\bibfield  {title} {\bibinfo {title} {Critical behavior in active lattice
  models of motility-induced phase separation},\ }\href
  {http://dx.doi.org/10.1140/epje/s10189-021-00058-1} {\bibfield  {journal}
  {\bibinfo  {journal} {Eur. Phys. J. E}\ }\textbf {\bibinfo {volume} {44}},\
  \bibinfo {pages} {53} (\bibinfo {year} {2021})}\BibitemShut {NoStop}%
\bibitem [{\citenamefont {Soto}\ and\ \citenamefont
  {Golestanian}(2014)}]{Soto2014}%
  \BibitemOpen
  \bibfield  {author} {\bibinfo {author} {\bibfnamefont {R.}~\bibnamefont
  {Soto}}\ and\ \bibinfo {author} {\bibfnamefont {R.}~\bibnamefont
  {Golestanian}},\ }\bibfield  {title} {\bibinfo {title} {Run-and-tumble
  dynamics in a crowded environment: Persistent exclusion process for
  swimmers},\ }\href {http://dx.doi.org/10.1103/PhysRevE.89.012706} {\bibfield
  {journal} {\bibinfo  {journal} {Phys. Rev. E}\ }\textbf {\bibinfo {volume}
  {89}},\ \bibinfo {pages} {012706} (\bibinfo {year} {2014})}\BibitemShut
  {NoStop}%
\bibitem [{\citenamefont {Whitelam}\ \emph {et~al.}(2018)\citenamefont
  {Whitelam}, \citenamefont {Klymko},\ and\ \citenamefont
  {Mandal}}]{Whitelam2018}%
  \BibitemOpen
  \bibfield  {author} {\bibinfo {author} {\bibfnamefont {S.}~\bibnamefont
  {Whitelam}}, \bibinfo {author} {\bibfnamefont {K.}~\bibnamefont {Klymko}},\
  and\ \bibinfo {author} {\bibfnamefont {D.}~\bibnamefont {Mandal}},\
  }\bibfield  {title} {\bibinfo {title} {Phase separation and large deviations
  of lattice active matter},\ }\href {http://dx.doi.org/10.1063/1.5023403}
  {\bibfield  {journal} {\bibinfo  {journal} {J. Chem. Phys.}\ }\textbf
  {\bibinfo {volume} {148}},\ \bibinfo {pages} {154902} (\bibinfo {year}
  {2018})}\BibitemShut {NoStop}%
\bibitem [{\citenamefont {Yao}\ and\ \citenamefont {Jack}(2025)}]{Yao2025}%
  \BibitemOpen
  \bibfield  {author} {\bibinfo {author} {\bibfnamefont {L.}~\bibnamefont
  {Yao}}\ and\ \bibinfo {author} {\bibfnamefont {R.~L.}\ \bibnamefont {Jack}},\
  }\bibfield  {title} {\bibinfo {title} {Interfacial and density fluctuations
  in a lattice model of motility-induced phase separation},\ }\href
  {http://dx.doi.org/10.1063/5.0253530} {\bibfield  {journal} {\bibinfo
  {journal} {J. Chem. Phys.}\ }\textbf {\bibinfo {volume} {162}},\ \bibinfo
  {pages} {114902} (\bibinfo {year} {2025})}\BibitemShut {NoStop}%
\bibitem [{\citenamefont {Solon}\ \emph {et~al.}(2015)\citenamefont {Solon},
  \citenamefont {Caussin}, \citenamefont {Bartolo}, \citenamefont {Chaté},\
  and\ \citenamefont {Tailleur}}]{Solon2015}%
  \BibitemOpen
  \bibfield  {author} {\bibinfo {author} {\bibfnamefont {A.~P.}\ \bibnamefont
  {Solon}}, \bibinfo {author} {\bibfnamefont {J.-B.}\ \bibnamefont {Caussin}},
  \bibinfo {author} {\bibfnamefont {D.}~\bibnamefont {Bartolo}}, \bibinfo
  {author} {\bibfnamefont {H.}~\bibnamefont {Chaté}},\ and\ \bibinfo {author}
  {\bibfnamefont {J.}~\bibnamefont {Tailleur}},\ }\bibfield  {title} {\bibinfo
  {title} {Pattern formation in flocking models: A hydrodynamic description},\
  }\href {http://dx.doi.org/10.1103/PhysRevE.92.062111} {\bibfield  {journal}
  {\bibinfo  {journal} {Phys. Rev. E}\ }\textbf {\bibinfo {volume} {92}},\
  \bibinfo {pages} {062111} (\bibinfo {year} {2015})}\BibitemShut {NoStop}%
\bibitem [{\citenamefont {Siebert}\ \emph {et~al.}(2018)\citenamefont
  {Siebert}, \citenamefont {Dittrich}, \citenamefont {Schmid}, \citenamefont
  {Binder}, \citenamefont {Speck},\ and\ \citenamefont {Virnau}}]{Siebert2018}%
  \BibitemOpen
  \bibfield  {author} {\bibinfo {author} {\bibfnamefont {J.~T.}\ \bibnamefont
  {Siebert}}, \bibinfo {author} {\bibfnamefont {F.}~\bibnamefont {Dittrich}},
  \bibinfo {author} {\bibfnamefont {F.}~\bibnamefont {Schmid}}, \bibinfo
  {author} {\bibfnamefont {K.}~\bibnamefont {Binder}}, \bibinfo {author}
  {\bibfnamefont {T.}~\bibnamefont {Speck}},\ and\ \bibinfo {author}
  {\bibfnamefont {P.}~\bibnamefont {Virnau}},\ }\bibfield  {title} {\bibinfo
  {title} {Critical behavior of active {Brownian} particles},\ }\href
  {http://dx.doi.org/10.1103/PhysRevE.98.030601} {\bibfield  {journal}
  {\bibinfo  {journal} {Phys. Rev. E}\ }\textbf {\bibinfo {volume} {98}},\
  \bibinfo {pages} {030601} (\bibinfo {year} {2018})}\BibitemShut {NoStop}%
\bibitem [{\citenamefont {Saha}\ \emph {et~al.}(2024)\citenamefont {Saha},
  \citenamefont {Banerjee},\ and\ \citenamefont {Mohanty}}]{Soumya2024}%
  \BibitemOpen
  \bibfield  {author} {\bibinfo {author} {\bibfnamefont {S.~K.}\ \bibnamefont
  {Saha}}, \bibinfo {author} {\bibfnamefont {A.}~\bibnamefont {Banerjee}},\
  and\ \bibinfo {author} {\bibfnamefont {P.~K.}\ \bibnamefont {Mohanty}},\
  }\bibfield  {title} {\bibinfo {title} {Site-percolation transition of
  run-and-tumble particles},\ }\href {https://doi.org/10.1039/D4SM00838C}
  {\bibfield  {journal} {\bibinfo  {journal} {Soft Matter}\ }\textbf {\bibinfo
  {volume} {20}},\ \bibinfo {pages} {9503} (\bibinfo {year}
  {2024})}\BibitemShut {NoStop}%
\bibitem [{\citenamefont {Partridge}\ and\ \citenamefont
  {Lee}(2019)}]{Partridge2019}%
  \BibitemOpen
  \bibfield  {author} {\bibinfo {author} {\bibfnamefont {B.}~\bibnamefont
  {Partridge}}\ and\ \bibinfo {author} {\bibfnamefont {C.~F.}\ \bibnamefont
  {Lee}},\ }\bibfield  {title} {\bibinfo {title} {Critical motility-induced
  phase separation belongs to the ising universality class},\ }\href
  {http://dx.doi.org/10.1103/PhysRevLett.123.068002} {\bibfield  {journal}
  {\bibinfo  {journal} {Phys. Rev. Lett.}\ }\textbf {\bibinfo {volume} {123}},\
  \bibinfo {pages} {068002} (\bibinfo {year} {2019})}\BibitemShut {NoStop}%
\bibitem [{\citenamefont {Maggi}\ \emph {et~al.}(2021)\citenamefont {Maggi},
  \citenamefont {Paoluzzi}, \citenamefont {Crisanti}, \citenamefont
  {Zaccarelli},\ and\ \citenamefont {Gnan}}]{Maggi2021}%
  \BibitemOpen
  \bibfield  {author} {\bibinfo {author} {\bibfnamefont {C.}~\bibnamefont
  {Maggi}}, \bibinfo {author} {\bibfnamefont {M.}~\bibnamefont {Paoluzzi}},
  \bibinfo {author} {\bibfnamefont {A.}~\bibnamefont {Crisanti}}, \bibinfo
  {author} {\bibfnamefont {E.}~\bibnamefont {Zaccarelli}},\ and\ \bibinfo
  {author} {\bibfnamefont {N.}~\bibnamefont {Gnan}},\ }\bibfield  {title}
  {\bibinfo {title} {Universality class of the motility-induced critical point
  in large scale off-lattice simulations of active particles},\ }\href
  {https://doi.org/10.1039/d0sm02162h} {\bibfield  {journal} {\bibinfo
  {journal} {J. Soft Matter}\ }\textbf {\bibinfo {volume} {17}},\ \bibinfo
  {pages} {3807} (\bibinfo {year} {2021})}\BibitemShut {NoStop}%
\bibitem [{\citenamefont {Sarkar}\ and\ \citenamefont {Basu}(2025)}]{Urna2025}%
  \BibitemOpen
  \bibfield  {author} {\bibinfo {author} {\bibfnamefont {R.}~\bibnamefont
  {Sarkar}}\ and\ \bibinfo {author} {\bibfnamefont {U.}~\bibnamefont {Basu}},\
  }\bibfield  {title} {\bibinfo {title} {Emergent short-range repulsion for
  attractively coupled active particles},\ }\href
  {https://doi.org/10.1039/D5SM00137D} {\bibfield  {journal} {\bibinfo
  {journal} {Soft Matter}\ }\textbf {\bibinfo {volume} {21}},\ \bibinfo {pages}
  {3595} (\bibinfo {year} {2025})}\BibitemShut {NoStop}%
\bibitem [{\citenamefont {Caprini}\ and\ \citenamefont
  {L\"{o}wen}(2023)}]{Caprini2023}%
  \BibitemOpen
  \bibfield  {author} {\bibinfo {author} {\bibfnamefont {L.}~\bibnamefont
  {Caprini}}\ and\ \bibinfo {author} {\bibfnamefont {H.}~\bibnamefont
  {L\"{o}wen}},\ }\bibfield  {title} {\bibinfo {title} {Flocking without
  alignment interactions in attractive active {Brownian} particles},\ }\href
  {http://dx.doi.org/10.1103/PhysRevLett.130.148202} {\bibfield  {journal}
  {\bibinfo  {journal} {Phys. Rev. Lett.}\ }\textbf {\bibinfo {volume} {130}},\
  \bibinfo {pages} {148202} (\bibinfo {year} {2023})}\BibitemShut {NoStop}%
\bibitem [{\citenamefont {Chakraborti}\ and\ \citenamefont
  {Zaburdaev}(2024)}]{Chakraborti2024}%
  \BibitemOpen
  \bibfield  {author} {\bibinfo {author} {\bibfnamefont {S.}~\bibnamefont
  {Chakraborti}}\ and\ \bibinfo {author} {\bibfnamefont {V.}~\bibnamefont
  {Zaburdaev}},\ }\bibfield  {title} {\bibinfo {title} {Transport in cellular
  aggregates described by fluctuating hydrodynamics},\ }\href
  {http://dx.doi.org/10.1103/PhysRevResearch.6.043064} {\bibfield  {journal}
  {\bibinfo  {journal} {Phys. Rev. Res.}\ }\textbf {\bibinfo {volume} {6}},\
  \bibinfo {pages} {043064} (\bibinfo {year} {2024})}\BibitemShut {NoStop}%
\bibitem [{\citenamefont {Shimamura}\ \emph {et~al.}()\citenamefont
  {Shimamura}, \citenamefont {Saito},\ and\ \citenamefont
  {Ishihara}}]{Sota2025}%
  \BibitemOpen
  \bibfield  {author} {\bibinfo {author} {\bibfnamefont {S.}~\bibnamefont
  {Shimamura}}, \bibinfo {author} {\bibfnamefont {N.}~\bibnamefont {Saito}},\
  and\ \bibinfo {author} {\bibfnamefont {S.}~\bibnamefont {Ishihara}},\
  }\href@noop {} {\bibinfo {title} {Attraction-induced cluster fragmentation
  and local alignment in active particle systems}},\ \Eprint
  {https://arxiv.org/abs/2505.19118} {arXiv:2505.19118} \BibitemShut {NoStop}%
\bibitem [{\citenamefont {Onsager}(1944)}]{Onsager1944}%
  \BibitemOpen
  \bibfield  {author} {\bibinfo {author} {\bibfnamefont {L.}~\bibnamefont
  {Onsager}},\ }\bibfield  {title} {\bibinfo {title} {Crystal statistics. i. a
  two-dimensional model with an order-disorder transition},\ }\href
  {https://doi.org/10.1103/physrev.65.117} {\bibfield  {journal} {\bibinfo
  {journal} {Phys. Rev.}\ }\textbf {\bibinfo {volume} {65}},\ \bibinfo {pages}
  {117} (\bibinfo {year} {1944})}\BibitemShut {NoStop}%
\bibitem [{\citenamefont {Gonnella}\ \emph {et~al.}(2015)\citenamefont
  {Gonnella}, \citenamefont {Marenduzzo}, \citenamefont {Suma},\ and\
  \citenamefont {Tiribocchi}}]{Gonnella2015}%
  \BibitemOpen
  \bibfield  {author} {\bibinfo {author} {\bibfnamefont {G.}~\bibnamefont
  {Gonnella}}, \bibinfo {author} {\bibfnamefont {D.}~\bibnamefont
  {Marenduzzo}}, \bibinfo {author} {\bibfnamefont {A.}~\bibnamefont {Suma}},\
  and\ \bibinfo {author} {\bibfnamefont {A.}~\bibnamefont {Tiribocchi}},\
  }\bibfield  {title} {\bibinfo {title} {Motility-induced phase separation and
  coarsening in active matter},\ }\href
  {https://doi.org/10.1016/j.crhy.2015.05.001} {\bibfield  {journal} {\bibinfo
  {journal} {Compt. Rend. Phys.}\ }\textbf {\bibinfo {volume} {16}},\ \bibinfo
  {pages} {316} (\bibinfo {year} {2015})}\BibitemShut {NoStop}%
\bibitem [{\citenamefont {Martin-Roca}\ \emph {et~al.}(2021)\citenamefont
  {Martin-Roca}, \citenamefont {Martinez}, \citenamefont {Alexander},
  \citenamefont {Diez}, \citenamefont {Aarts}, \citenamefont {Alarcon},
  \citenamefont {Ram{\'{\i}}rez},\ and\ \citenamefont
  {Valeriani}}]{MartinRoca2021}%
  \BibitemOpen
  \bibfield  {author} {\bibinfo {author} {\bibfnamefont {J.}~\bibnamefont
  {Martin-Roca}}, \bibinfo {author} {\bibfnamefont {R.}~\bibnamefont
  {Martinez}}, \bibinfo {author} {\bibfnamefont {L.~C.}\ \bibnamefont
  {Alexander}}, \bibinfo {author} {\bibfnamefont {A.~L.}\ \bibnamefont {Diez}},
  \bibinfo {author} {\bibfnamefont {D.~G. A.~L.}\ \bibnamefont {Aarts}},
  \bibinfo {author} {\bibfnamefont {F.}~\bibnamefont {Alarcon}}, \bibinfo
  {author} {\bibfnamefont {J.}~\bibnamefont {Ram{\'{\i}}rez}},\ and\ \bibinfo
  {author} {\bibfnamefont {C.}~\bibnamefont {Valeriani}},\ }\bibfield  {title}
  {\bibinfo {title} {Characterization of {MIPS} in a suspension of repulsive
  active {Brownian} particles through dynamical features},\ }\href
  {https://doi.org/10.1063/5.0040141} {\bibfield  {journal} {\bibinfo
  {journal} {J. Chem. Phys.}\ }\textbf {\bibinfo {volume} {154}},\ \bibinfo
  {pages} {174901} (\bibinfo {year} {2021})}\BibitemShut {NoStop}%
\bibitem [{\citenamefont {Rovere}\ \emph {et~al.}(1993)\citenamefont {Rovere},
  \citenamefont {Nielaba},\ and\ \citenamefont {Binder}}]{Rovere1993}%
  \BibitemOpen
  \bibfield  {author} {\bibinfo {author} {\bibfnamefont {M.}~\bibnamefont
  {Rovere}}, \bibinfo {author} {\bibfnamefont {P.}~\bibnamefont {Nielaba}},\
  and\ \bibinfo {author} {\bibfnamefont {K.}~\bibnamefont {Binder}},\
  }\bibfield  {title} {\bibinfo {title} {Simulation studies of gas-liquid
  transitions in two dimensions via a subsystem-block-density distribution
  analysis},\ }\href {https://doi.org/10.1007/bf02198158} {\bibfield  {journal}
  {\bibinfo  {journal} {Z. Phys. B}\ }\textbf {\bibinfo {volume} {90}},\
  \bibinfo {pages} {215} (\bibinfo {year} {1993})}\BibitemShut {NoStop}%
\bibitem [{\citenamefont {Binder}(1981)}]{Binder1981}%
  \BibitemOpen
  \bibfield  {author} {\bibinfo {author} {\bibfnamefont {K.}~\bibnamefont
  {Binder}},\ }\bibfield  {title} {\bibinfo {title} {Finite size scaling
  analysis of ising model block distribution functions},\ }\href
  {https://doi.org/10.1007/bf01293604} {\bibfield  {journal} {\bibinfo
  {journal} {Z. Phys. B}\ }\textbf {\bibinfo {volume} {43}},\ \bibinfo {pages}
  {119} (\bibinfo {year} {1981})}\BibitemShut {NoStop}%
\bibitem [{\citenamefont {Bai}\ and\ \citenamefont {Breen}(2008)}]{BaiBreen}%
  \BibitemOpen
  \bibfield  {author} {\bibinfo {author} {\bibfnamefont {L.}~\bibnamefont
  {Bai}}\ and\ \bibinfo {author} {\bibfnamefont {D.}~\bibnamefont {Breen}},\
  }\bibfield  {title} {\bibinfo {title} {Calculating center of mass in an
  unbounded {2D} environment},\ }\href
  {https://doi.org/10.1080/2151237X.2008.10129266} {\bibfield  {journal}
  {\bibinfo  {journal} {J. Graph. Tools}\ }\textbf {\bibinfo {volume} {13}},\
  \bibinfo {pages} {53} (\bibinfo {year} {2008})}\BibitemShut {NoStop}%
\bibitem [{\citenamefont {Stauffer}(1979)}]{Stauffer1979}%
  \BibitemOpen
  \bibfield  {author} {\bibinfo {author} {\bibfnamefont {D.}~\bibnamefont
  {Stauffer}},\ }\bibfield  {title} {\bibinfo {title} {Scaling theory of
  percolation clusters},\ }\href {https://doi.org/10.1016/0370-1573(79)90060-7}
  {\bibfield  {journal} {\bibinfo  {journal} {Phys. Rep.}\ }\textbf {\bibinfo
  {volume} {54}},\ \bibinfo {pages} {1} (\bibinfo {year} {1979})}\BibitemShut
  {NoStop}%
\bibitem [{\citenamefont {Essam}(1980)}]{Essam1980}%
  \BibitemOpen
  \bibfield  {author} {\bibinfo {author} {\bibfnamefont {J.~W.}\ \bibnamefont
  {Essam}},\ }\bibfield  {title} {\bibinfo {title} {Percolation theory},\
  }\href {https://doi.org/10.1088/0034-4885/43/7/001} {\bibfield  {journal}
  {\bibinfo  {journal} {Rep. Prog. Phys.}\ }\textbf {\bibinfo {volume} {43}},\
  \bibinfo {pages} {833} (\bibinfo {year} {1980})}\BibitemShut {NoStop}%
\bibitem [{\citenamefont {Coniglio}\ and\ \citenamefont
  {Klein}(1980)}]{Coniglio1980}%
  \BibitemOpen
  \bibfield  {author} {\bibinfo {author} {\bibfnamefont {A.}~\bibnamefont
  {Coniglio}}\ and\ \bibinfo {author} {\bibfnamefont {W.}~\bibnamefont
  {Klein}},\ }\bibfield  {title} {\bibinfo {title} {Clusters and ising critical
  droplets: a renormalisation group approach},\ }\href
  {https://doi.org/10.1088/0305-4470/13/8/025} {\bibfield  {journal} {\bibinfo
  {journal} {J. Phys. A: Math. Gen.}\ }\textbf {\bibinfo {volume} {13}},\
  \bibinfo {pages} {2775} (\bibinfo {year} {1980})}\BibitemShut {NoStop}%
\bibitem [{\citenamefont {Coniglio}(2001)}]{Coniglio2001}%
  \BibitemOpen
  \bibfield  {author} {\bibinfo {author} {\bibfnamefont {A.}~\bibnamefont
  {Coniglio}},\ }\bibfield  {title} {\bibinfo {title} {Percolation and critical
  points},\ }\href {https://doi.org/10.1088/0953-8984/13/41/301} {\bibfield
  {journal} {\bibinfo  {journal} {J. Phys.: Condens. Matter}\ }\textbf
  {\bibinfo {volume} {13}},\ \bibinfo {pages} {9039} (\bibinfo {year}
  {2001})}\BibitemShut {NoStop}%
\bibitem [{\citenamefont {Fortunato}(2002)}]{Fortunato}%
  \BibitemOpen
  \bibfield  {author} {\bibinfo {author} {\bibfnamefont {S.}~\bibnamefont
  {Fortunato}},\ }\bibfield  {title} {\bibinfo {title} {Site percolation and
  phase transitions in two dimensions},\ }\href
  {https://doi.org/10.1103/PhysRevB.66.054107} {\bibfield  {journal} {\bibinfo
  {journal} {Phys. Rev. B}\ }\textbf {\bibinfo {volume} {66}},\ \bibinfo
  {pages} {054107} (\bibinfo {year} {2002})}\BibitemShut {NoStop}%
\bibitem [{\citenamefont {Janke}\ and\ \citenamefont
  {Schakel}(2005)}]{Janke2005}%
  \BibitemOpen
  \bibfield  {author} {\bibinfo {author} {\bibfnamefont {W.}~\bibnamefont
  {Janke}}\ and\ \bibinfo {author} {\bibfnamefont {A.~M.~J.}\ \bibnamefont
  {Schakel}},\ }\bibfield  {title} {\bibinfo {title} {Fractal structure of spin
  clusters and domain walls in the two-dimensional ising model},\ }\href
  {http://dx.doi.org/10.1103/PhysRevE.71.036703} {\bibfield  {journal}
  {\bibinfo  {journal} {Phys. Rev. E}\ }\textbf {\bibinfo {volume} {71}},\
  \bibinfo {pages} {036703} (\bibinfo {year} {2005})}\BibitemShut {NoStop}%
\bibitem [{\citenamefont {Stella}\ and\ \citenamefont
  {Vanderzande}(1989)}]{Stella1989}%
  \BibitemOpen
  \bibfield  {author} {\bibinfo {author} {\bibfnamefont {A.~L.}\ \bibnamefont
  {Stella}}\ and\ \bibinfo {author} {\bibfnamefont {C.}~\bibnamefont
  {Vanderzande}},\ }\bibfield  {title} {\bibinfo {title} {Scaling and fractal
  dimension of {Ising} clusters at the {$d=2$} critical point},\ }\href
  {https://doi.org/10.1103/physrevlett.62.1067} {\bibfield  {journal} {\bibinfo
   {journal} {Phys. Rev. Lett.}\ }\textbf {\bibinfo {volume} {62}},\ \bibinfo
  {pages} {1067} (\bibinfo {year} {1989})}\BibitemShut {NoStop}%
\bibitem [{\citenamefont {Banerjee}\ \emph {et~al.}(2025)\citenamefont
  {Banerjee}, \citenamefont {Jana},\ and\ \citenamefont {Mohanty}}]{Aikya2025}%
  \BibitemOpen
  \bibfield  {author} {\bibinfo {author} {\bibfnamefont {A.}~\bibnamefont
  {Banerjee}}, \bibinfo {author} {\bibfnamefont {P.}~\bibnamefont {Jana}},\
  and\ \bibinfo {author} {\bibfnamefont {P.~K.}\ \bibnamefont {Mohanty}},\
  }\bibfield  {title} {\bibinfo {title} {Geometric percolation of spins and
  spin dipoles in the {Ashkin-Teller} model},\ }\href
  {https://doi.org/10.1103/PhysRevB.111.014403} {\bibfield  {journal} {\bibinfo
   {journal} {Phys. Rev. B}\ }\textbf {\bibinfo {volume} {111}},\ \bibinfo
  {pages} {014403} (\bibinfo {year} {2025})}\BibitemShut {NoStop}%
\bibitem [{\citenamefont {Stauffer}\ and\ \citenamefont
  {Aharony}(1994)}]{Stauffer1994}%
  \BibitemOpen
  \bibfield  {author} {\bibinfo {author} {\bibfnamefont {D.}~\bibnamefont
  {Stauffer}}\ and\ \bibinfo {author} {\bibfnamefont {A.}~\bibnamefont
  {Aharony}},\ }\href@noop {} {\emph {\bibinfo {title} {Introduction to
  {Percolation Theory}}}},\ \bibinfo {edition} {2nd}\ ed.\ (\bibinfo
  {publisher} {Taylor \& Francis},\ \bibinfo {address} {London},\ \bibinfo
  {year} {1994})\BibitemShut {NoStop}%
\bibitem [{sup()}]{supp}%
  \BibitemOpen
  \href@noop {} {\bibinfo {title} {{See Supplemental Material for details of
  the dynamics, simulation and estimation of critical points and exponents of
  percolation and phase separation transition, in support of our
  claims.}}}\BibitemShut {Stop}%
\bibitem [{\citenamefont {Saberi}\ and\ \citenamefont
  {Dashti-Naserabadi}(2010)}]{Saberi2010}%
  \BibitemOpen
  \bibfield  {author} {\bibinfo {author} {\bibfnamefont {A.~A.}\ \bibnamefont
  {Saberi}}\ and\ \bibinfo {author} {\bibfnamefont {H.}~\bibnamefont
  {Dashti-Naserabadi}},\ }\bibfield  {title} {\bibinfo {title}
  {Three-dimensional {Ising} model, percolation theory and conformal
  invariance},\ }\href {http://dx.doi.org/10.1209/0295-5075/92/67005}
  {\bibfield  {journal} {\bibinfo  {journal} {Europhys. Lett.}\ }\textbf
  {\bibinfo {volume} {92}},\ \bibinfo {pages} {67005} (\bibinfo {year}
  {2010})}\BibitemShut {NoStop}%
\bibitem [{\citenamefont {Rovere}\ \emph {et~al.}(1990)\citenamefont {Rovere},
  \citenamefont {Heermann},\ and\ \citenamefont {Binder}}]{Rovere1990}%
  \BibitemOpen
  \bibfield  {author} {\bibinfo {author} {\bibfnamefont {M.}~\bibnamefont
  {Rovere}}, \bibinfo {author} {\bibfnamefont {D.~W.}\ \bibnamefont
  {Heermann}},\ and\ \bibinfo {author} {\bibfnamefont {K.}~\bibnamefont
  {Binder}},\ }\bibfield  {title} {\bibinfo {title} {The gas-liquid transition
  of the two-dimensional {Lennard-Jones} fluid},\ }\href
  {https://doi.org/10.1088/0953-8984/2/33/013} {\bibfield  {journal} {\bibinfo
  {journal} {J. Phys.: Condens. Matter}\ }\textbf {\bibinfo {volume} {2}},\
  \bibinfo {pages} {7009} (\bibinfo {year} {1990})}\BibitemShut {NoStop}%
\bibitem [{\citenamefont {O’Byrne}\ \emph {et~al.}(2022)\citenamefont
  {O’Byrne}, \citenamefont {Kafri}, \citenamefont {Tailleur},\ and\
  \citenamefont {van Wijland}}]{OByrne2022}%
  \BibitemOpen
  \bibfield  {author} {\bibinfo {author} {\bibfnamefont {J.}~\bibnamefont
  {O’Byrne}}, \bibinfo {author} {\bibfnamefont {Y.}~\bibnamefont {Kafri}},
  \bibinfo {author} {\bibfnamefont {J.}~\bibnamefont {Tailleur}},\ and\
  \bibinfo {author} {\bibfnamefont {F.}~\bibnamefont {van Wijland}},\
  }\bibfield  {title} {\bibinfo {title} {Time irreversibility in active matter,
  from micro to macro},\ }\href {https://doi.org/10.1038/s42254-021-00406-2}
  {\bibfield  {journal} {\bibinfo  {journal} {Nat. Rev. Phys.}\ }\textbf
  {\bibinfo {volume} {4}},\ \bibinfo {pages} {167} (\bibinfo {year}
  {2022})}\BibitemShut {NoStop}%
\bibitem [{\citenamefont {Pelissetto}\ and\ \citenamefont
  {Vicari}(2000)}]{Vicari}%
  \BibitemOpen
  \bibfield  {author} {\bibinfo {author} {\bibfnamefont {A.}~\bibnamefont
  {Pelissetto}}\ and\ \bibinfo {author} {\bibfnamefont {E.}~\bibnamefont
  {Vicari}},\ }\bibfield  {title} {\bibinfo {title} {Critical phenomena and
  renormalization-group theory},\ }\href
  {https://doi.org/10.1103/PhysRevB.62.6393} {\bibfield  {journal} {\bibinfo
  {journal} {Phys. Rev. B}\ }\textbf {\bibinfo {volume} {62}},\ \bibinfo
  {pages} {6393} (\bibinfo {year} {2000})}\BibitemShut {NoStop}%
\bibitem [{\citenamefont {Mukherjee}\ and\ \citenamefont
  {Mohanty}(2023)}]{Mukherjee2023}%
  \BibitemOpen
  \bibfield  {author} {\bibinfo {author} {\bibfnamefont {I.}~\bibnamefont
  {Mukherjee}}\ and\ \bibinfo {author} {\bibfnamefont {P.~K.}\ \bibnamefont
  {Mohanty}},\ }\bibfield  {title} {\bibinfo {title} {Hidden superuniversality
  in systems with continuous variation of critical exponents},\ }\href
  {http://dx.doi.org/10.1103/PhysRevB.108.174417} {\bibfield  {journal}
  {\bibinfo  {journal} {Phys. Rev. B}\ }\textbf {\bibinfo {volume} {108}},\
  \bibinfo {pages} {174417} (\bibinfo {year} {2023})}\BibitemShut {NoStop}%
\end{thebibliography}%
	
	\ifarXiv
	\foreach \x in {1,...,\numbersupplementpages}
	{
		\clearpage
		\includepdf[pages={\x}]{\supplementfilename}
	}
	\fi
\end{document}